\documentclass[twocolumn]{aastex631}
\usepackage{changepage}
\usepackage[flushleft]{threeparttable}
\usepackage{rotating,booktabs}

\newcommand{\ewmgii}{$W_{r}^{2796+2803}$}
\def\ewmgiia{W$_{r}^{2796}$}
\newcommand{\kms}{km~s$^{-1}$}  
\newcommand{\sqcm}{cm$^{-2}$}  
\newcommand{\lya}{Ly$\alpha$}

\newcommand{\HI}{\mbox{H\,{\sc i}}}

\newcommand{\Msun}{$\rm M_{\odot}$} 
\newcommand{\mvir}{$M_{\rm 500}$} 
\newcommand{\rvir}{$R_{\rm 500}$} 
\newcommand{\rhocl}{$\rho_{\rm cl}$} 
\newcommand{\zcl}{$z_{\rm cl}$} 
\newcommand{\nrhocl}{$\rho_{\rm cl}/R_{\rm 500}$}

\newcommand{\MgI}{\mbox{Mg\,{\sc i}}}
\newcommand{\MgII}{\mbox{Mg\,{\sc ii}}}
\newcommand{\FeII}{\mbox{Fe\,{\sc ii}}}
\def\mgiiab{Mg\,{\sc ii}$\lambda\lambda$2796,2803~}
\def\feiia{Fe\,{\sc ii}$\lambda$2600~} 
\def\mgiia{Mg~{\sc ii}$\lambda$2796~}

\def\mgii{Mg~{\sc ii}~} 
\def\feii{Fe~{\sc ii}~} 

\def\feii{Fe~{\sc ii}~}

\def\civ{C~{\sc iv}~}

\graphicspath{{./}{figures/}}

\begin{document}

\title{Discovery of a cool, metal-rich gas reservoir in the outskirts of $z\approx0.5$ clusters}
%\footnote{Released on March, 1st, 2021}}

%[0000-0002-0786-7307]
\author{Sapna Mishra}
\affiliation{Inter-University Centre for Astronomy and Astrophysics (IUCAA), Post Bag 4, Ganeshkhind, Pune 411 007, India}

\author{Sowgat Muzahid}
\affiliation{Inter-University Centre for Astronomy and Astrophysics (IUCAA), Post Bag 4, Ganeshkhind, Pune 411 007, India}
\affiliation{Leibniz-Institute for Astrophysics Potsdam (AIP), An der Sternwarte 16, D-14482 Potsdam, Germany} 

\begin{abstract}

We built the first-ever statistically significant sample of $\approx80,000$ background quasar -- foreground cluster pairs to study the cool, metal-rich gas in the outskirts ($>R_{500}$) of $z\approx0.5$ clusters with a median mass of $\approx 10^{14.2}$~\Msun. The sample was obtained by cross-matching the SDSS cluster catalog of \citet{2015ApJ...807..178W} and SDSS quasar catalog of \citet{2020ApJS..250....8L}. The median impact parameter (\rhocl) of the clusters from the quasar sightlines is $2.4$~Mpc (median \nrhocl~$=3.6$). A strong \MgII, along with marginal \FeII, absorption is detected in the mean and median stacked spectra of the quasars with total \MgII\ rest-frame equivalent width (\ewmgii) of $0.034\pm 0.005$~\AA\ (7$\sigma$) and  $0.010\pm 0.003$~\AA\ (3$\sigma$),  respectively. The \ewmgii shows a declining trend with increasing \rhocl\ and \nrhocl, but does not show any significant trend with mass ($M_{500}$) or redshift (\zcl) within the small \mvir\ and \zcl\ ranges probed here. The \MgII\ absorption signal and the trends persist even if we exclude the quasar-cluster pairs where the background quasars may be probing the circum-galactic medium (CGM) of bright galaxies with impact parameters $<300$~kpc. The \MgII\ (and \FeII) absorption reported here is the first detection of its kind. It indicates the presence of a cool, metal-rich gas reservoir surrounding galaxy clusters out to several \rvir. We suggest that the metal-rich gas in the cluster outskirts arise from stripped materials and that gas stripping may be important out to large clustocentric distances ($>3R_{500}$).     

\end{abstract}

\keywords{galaxies: clusters: general -- galaxies: clusters: intracluster medium -- (galaxies:) intergalactic medium -- (galaxies:) quasars: absorption lines}

\section{Introduction} 
\label{sec:intro}

Galaxy clusters, being the largest structures in the universe, serve as unique environments for studying galaxy evolution. They are testbeds for cosmological quandaries such as the missing baryons, properties of dark matter and dark energy, and enrichment of the intergalactic medium \citep[IGM;][]{Gonzalez2007,Allen2008,Vikhlinin2009}. Estimating how baryons are distributed among different components and phases in clusters is key for understanding the roles played by different physical processes such as starburst winds, supernovae-/AGN-feedback, galaxy mergers, and tidal/ram-pressure stripping in driving galaxy evolution in such complex environments. The majority of the baryons in clusters reside in the inter-cluster medium (ICM) in an X-ray emitting, hot ($\sim10^{7-8}$~K) phase \citep{Ettori2003}. Consequently, the central regions of clusters are well studied via X-ray (and GHz) observations \citep[e.g.,][]{Bleem15,Liu2021}. In addition, sensitive X-ray observations of a handful of nearby clusters have characterized the properties such as the surface brightness, metallicity, and gas fraction out to the virial radii  \citep[$R_{\rm vir}$;][]{Simionescu2011,Urban2014}. However, owing to the significantly lower density and temperature, the outskirts (i.e., $1-5\times R_{500}$\footnote{$R_{500}$ is the radius within which the mean mass density of a cluster is 500 times the critical density of the universe.})  or the circum-cluster medium (CCM) of galaxy clusters are not well explored. This is primarily because of the lack of sensitive diagnostics in X-rays to probe the cool/warm-hot gas ($\sim10^{4-6}$~K) that prevails in the CCM.

With the advent of high resolution cosmological hydrodynamical simulations, cluster outskirts have been recognised as an important environment for studying cluster astrophysics and cosmology \citep{Nagai2007,Lau2015,Bahe2017a}. Being the interface between clusters and the IGM, the CCM acts as ``melting pots'' where infalling metal-poor gas from the IGM mixes with metal-rich gas when galaxies and groups of galaxies are stripped of via ram pressure and tidal forces. Thus, it is a new frontier for understanding gas flows in and around clusters, and the environmental processes that drive galaxy evolution in clusters \citep[see][and references therein]{walker2019}. Using hydrodynamical simulations, \citet{Emerick2015} showed that the majority of the cool/warm-hot gas in cluster outskirts is linked with infalling material through IGM filaments. This, however, changes in the inner parts, where the stripping of galaxies becomes important. They found that the cool/warm-hot gas fraction increases with radii and become comparable to the hot gas fraction at $R\gtrsim 3R_{\rm vir}$. The latter study by \citet{Butsky2019} also reported qualitatively similar results. They showed that the ICM exhibits an increasingly multiphase nature at large clustocentric distances (i.e., in the outskirts). The metallicity of the cool/warm-hot gas in the outskirts shows a large scatter compared to the hot gas in the cores, owing to the poor mixing of IGM gas with stripped of galactic materials. Finally, they found signatures of CGM stripping of cluster galaxies during the early stages of infall at distances as far as $4R_{\rm vir}$.

There are only a handful of observational studies in the literature focused on the CCM \citep[e.g.,][]{lopez2008,yoon2012,Muzahid2017,Yoon2017,Burchett2018}. The study of \citet{yoon2012} showed that \lya\ absorbers are preferentially located in the outskirts of Virgo clusters and are associated with substructures traced by \HI\ 21-cm emission. \citet{Muzahid2017} reported the presence of partial Lyman limit systems (LLS) with $N(\HI)\gtrsim10^{16.5}$~\sqcm\ in the outskirts of 3/3 clusters they studied. Detailed ionization modelling of those absorption systems revealed the presence of metal-enriched gas with metallicities of $\rm [X/H] > -1$ \citep[]{Pradeep2019}. \citet{Burchett2018} studied the outskirts of 7 clusters but found no significant \HI~ absorption. Using ground-based observations, \citet{lopez2008} reported an overabundance of strong \MgII\ absorption with rest-frame equivalent width (REW) of \mgiia (\ewmgiia) $>2$~\AA\ compared to the ``field'' \MgII\ population \citep[see also][]{lee2021}. By cross-matching catalogs of \MgII\ absorbers and clusters from Dark Energy Spectroscopic Instrument (DESI) survey, \citet{anand2022cool} recently reported a covering fraction of 1--5\% within $R_{500}$ for \ewmgiia $>0.4$~\AA. From a lack of correlations  between the absorbers and the properties of nearest cluster galaxies (within $R_{200}$), and the fact that the median absorber-galaxy separation is $\approx200$~kpc, they concluded that \MgII\ absorption stems from stripped ISM and/or from satellite galaxies.

All of these aforementioned studies dealt either with small sample sizes and/or limited by the sensitivity to detect individual absorbers. To investigate the extent of cool, metal-enriched gas and their dependence on the cluster properties using a statistically significant sample, here we employ spectral stacking technique to a large number ($\approx$~80,000) of quasar-cluster pairs. The paper is organised as follows: In Section~\ref{sec:data} we describe the quasar and cluster samples used in this study, followed by the details of the construction of the quasar-cluster pairs, and continuum normalization of the SDSS spectra.  The main results are presented in Section~\ref{sec:result} followed by a discussion in Section~\ref{sec:discussion}. We adopted a flat $\Lambda$CDM cosmology with H$_{0} =$ 71 km s$^{-1}$ Mpc$^{-1}$ , $\Omega_{\rm M} =$ 0.3, and $\Omega_{\Lambda} =$ 0.7.

\section{Samples}   
\label{sec:data} 

\subsection{Cluster sample}  
\label{subsec:cluster-sample} 

We use the Sloan Digital Sky Survey (SDSS) galaxy cluster catalog of \citet[][hereafter WH15]{2015ApJ...807..178W}, consisting of 158,103 clusters. The WH15 improves upon the catalog of \citet[][hereafter WHL12]{2012ApJS..199...34W} and provides the largest number of clusters with spectroscopic redshifts. The masses of the clusters in WH15 are calculated using a scaling relation derived from a sample of 1191 clusters with well-measured masses with $M_{500} > 0.3 \times10^{14}$~\Msun\ and $0.05 < z_{\rm cl} < 0.75$. Therefore, we start with the 156,139 clusters within this mass and redshift ranges. Our requirement of having spectroscopic redshifts reduces the number to 119,653 clusters. To eliminate repeated entries with somewhat different redshifts and sky positions within these clusters, we flag the clusters that met the following criteria: (i) two clusters with a velocity offset $<$ 1000~\kms\ and (ii) physical separation is less than the sum of their $R_{500}$ values\footnote{Note that we recalculated the $R_{500}$ values using the redshift and $M_{500}$ values from the catalog for our adopted cosmology.}. We only consider the most massive one if two or more clusters satisfy these two conditions. This results in 118,000 unique clusters. 

While no formal completeness estimate is presented in WH15, the WHL12 catalog, which forms the bulk of the WH15 sample, is $>95\%$ complete for $M_{\rm 500} >10^{14} ~\mathrm{M}_{\odot}$ and $0.05<$~\zcl~$<0.42$. Note that all the clusters in our sample have \zcl~$>0.42$ (Section~\ref{subsec:qso-cluster-pairs}) for which the completeness of the cluster catalog is not characterized. However, 88\% of the clusters in our sample have $M_{500} > 10^{14}$~\Msun, and the results presented here are consistent if the clusters with $M_{500} < 10^{14}$~\Msun\ are excluded.

\subsection{Quasar sample}  
\label{subsec:qso-sample} 

We use the SDSS DR16 quasar catalog of \citet{2020ApJS..250....8L} to cross-match with the cluster catalog. This catalog contains a total of 750,414 quasars. We exclude the spectra with median signal-to-noise (SNR) of $< 5$, reducing the number to 291,956. We also exclude the spectra with broad absorption lines (BALs) with a non-zero ``BALnicity index'' (BI) as provided in the catalog of \citet{2020ApJS..250....8L}. This further reduces our sample to 278,580 quasars which is then used to cross-match with the cluster catalog described above. The spectroscopic observations of these quasars are performed using SDSS and Baryon Oscillation Spectroscopic Survey (BOSS) spectrographs. The spectra obtained with SDSS cover a spectral range between 3800$-$9200 \AA\ with spectral resolution ($R$)  ranging from 1850$-$2200. The BOSS spectra cover a spectral range from 3600$-$10,000 \AA\ with $R\approx 1300-3000$.

\subsection{Quasar-cluster pairs}  
\label{subsec:qso-cluster-pairs}
We cross-match the catalog of 278,580 background quasars with the 118,000 galaxy clusters. We impose the following selection criteria for our quasar-cluster pairs: 

(i) The cluster must have a spectroscopic redshift (\zcl) for which the \mgii\ doublet should fall in the wavelength range 4000$-$9000\AA, since the throughputs of SDSS and BOSS drop below 10\% beyond this range, \citep{Smee2013} leading to poor spectral SNR.  

\smallskip 
(ii) The cluster redshift (\zcl) is such that the \MgII\ doublet always falls on the redward of the \lya\ emission of the background quasar. This is to avoid contamination from the \lya-forest absorption. 

\smallskip 
(iii) The line-of-sight velocity offset between the quasar and cluster redshifts is $>5000$~\kms\ to minimize possible contamination due to absorption intrinsic to the background quasar/host-galaxy \citep[e.g.,][]{Muzahid2013}. 

\smallskip 
(iv) The projected separation between the foreground cluster and the background quasar at the redshift of the cluster (\rhocl) should lie between 1$-$5 $\times~ R_{500}$, suitable to probe the CCM. 

%---------------------------------Fig 1
\begin{figure}
\begin{center} 
\includegraphics[width=0.5\textwidth]{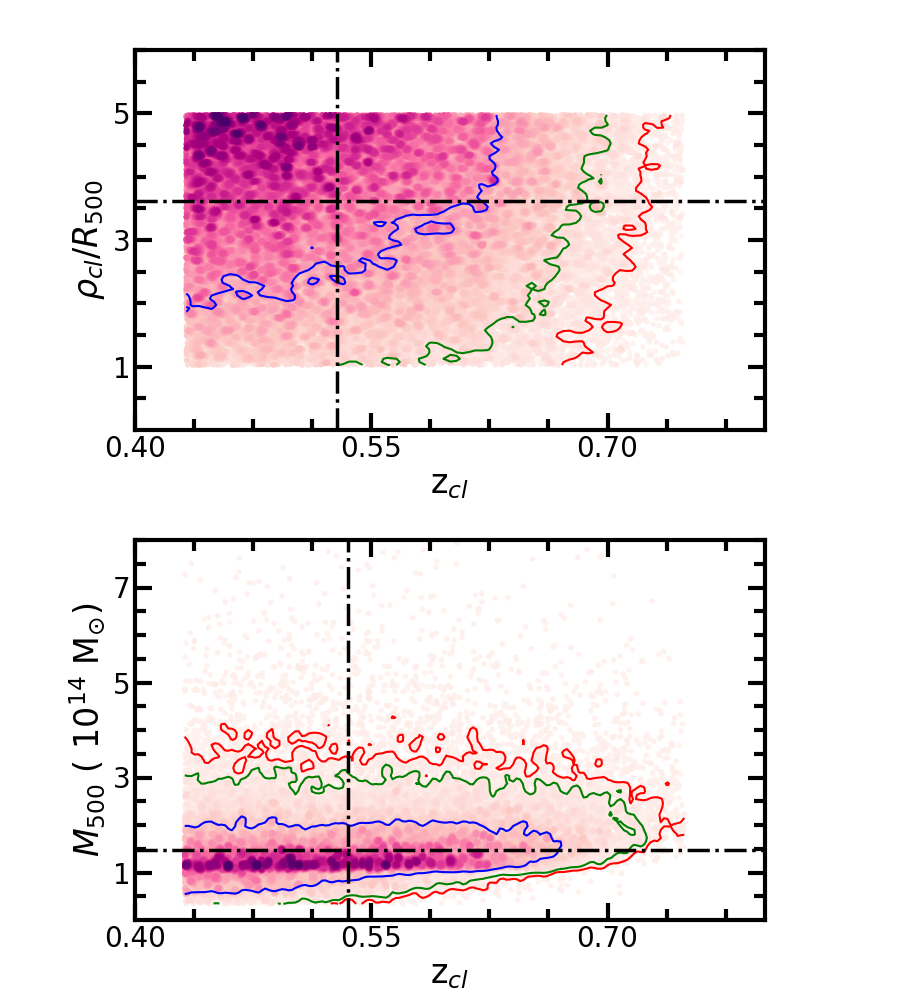}
     \caption{$M_{500}$ versus \zcl\ for the 37,632 unique clusters ({\em bottom}), and \nrhocl\ versus \zcl\ for the 79,485 quasar-cluster pairs ({\em top}). The density maps, along with the 68, 95, and 99.9 percentile contours are shown in blue, green, and red. The black dashed horizontal and vertical lines show the median value of the abscissa and ordinate parameters, respectively.}
     \label{fig:sample} 
\end{center} 
\end{figure}
%----------------------------

%---------------------------------Fig 2
\begin{figure*}
\begin{center} 
      \includegraphics[width=0.95\textwidth]{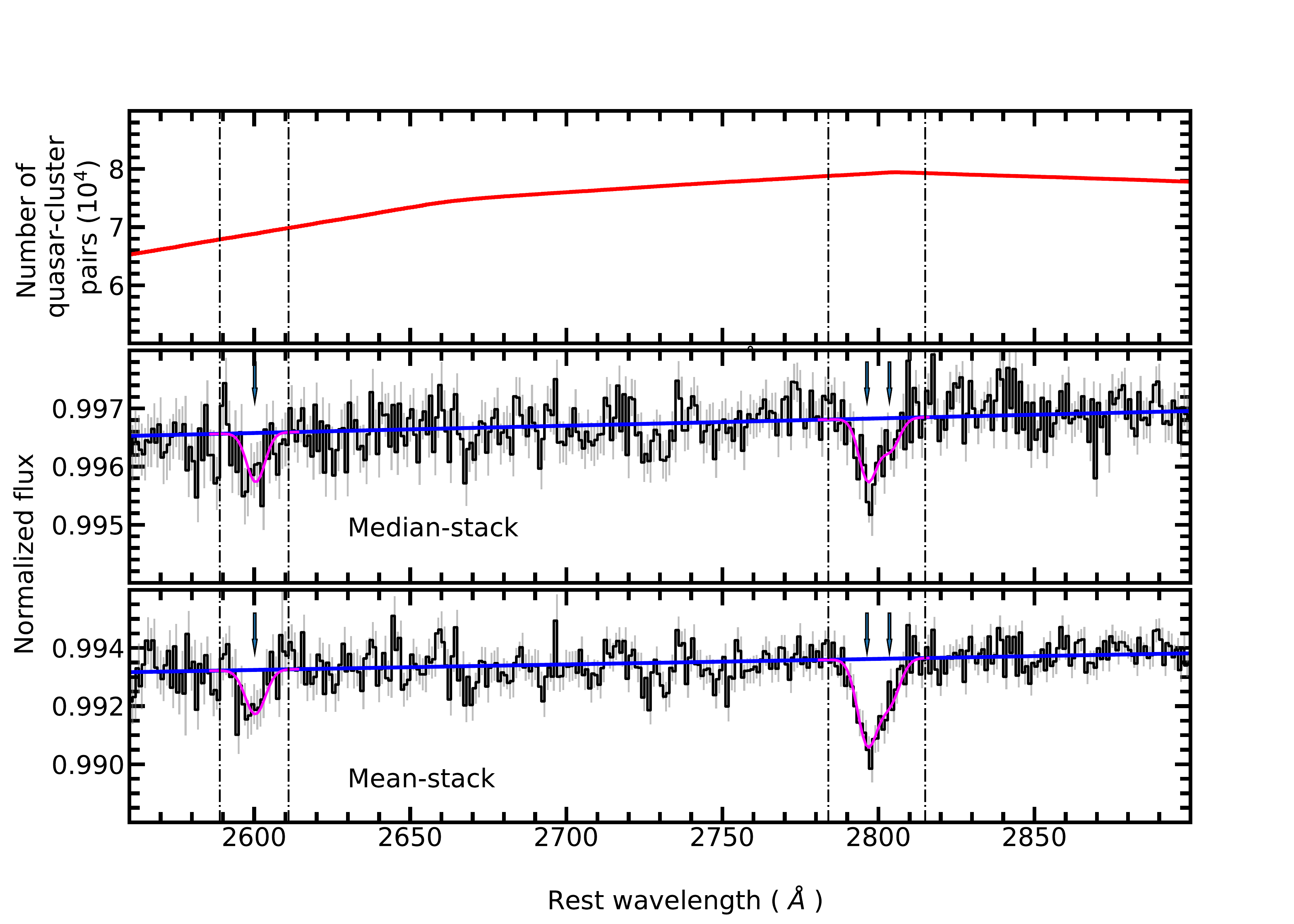}
     \caption{{\em Bottom:} The mean composite spectrum of the background quasars at the rest-frame of the clusters for the full sample. The grey $1 \sigma$ error-bars are estimated from 200 bootstrap realizations. The solid blue line shows the best-fit pseudo continuum. The best-fit \mgii and \feii absorption profiles are shown in smooth magenta curves. The line centroids are shown by the arrows. {\em Middle:} Same as the bottom panel but for the median stack. \emph{Top:} Number of quasar-cluster pairs contributing to the stacked spectra as a function of rest-frame wavelength. The four dashed horizontal lines two each around the \feii and \mgii lines in each panel, respectively, correspond to $[2589, 2611]$~\AA\ and $[2784, 2815]$~\AA\ wavelength ranges within which the REWs are calculated.} 
     \label{fig:stackprofiles} 
\end{center} 
\end{figure*}
%--------------------------------------

The aforementioned conditions yield 81,045 quasar-cluster pairs with 64,339 unique quasars and 38,001 unique clusters. In the subsequent analysis, we exclude 887 spectra contributing to 1104 quasar-cluster pairs in which we noticed spectral gaps in our regions of interest. Next, even after excluding the quasars with non-zero BI \citep[]{2020ApJS..250....8L}, we found strong BAL features in 326 quasar spectra (284 \civ BAL quasars and 42 \mgii BAL quasars), accounting for a total of 456 quasar-cluster pairs\footnote{We use a method similar to \citet{Mishra2021} to search for \civ\ and \mgii\ BAL features.}. After excluding those 456 pairs, our final sample has  79,485 quasar-cluster pairs with 63,126 and 37,632 unique quasars and clusters, respectively.

In Fig.~\ref{fig:sample}, we show the scatter plots of $M_{500}$ vs \zcl\ for the 37,632 unique clusters (bottom panel) and \nrhocl\ vs \zcl\ for the 79,485 quasar-cluster pairs (top panel). The cluster redshifts range from $0.43$ to $0.75$ with a median of $0.54$. The $M_{500}$ ranges from $(0.3-12.5) \times10^{14}$~\Msun\ with a median of $1.5\times10^{14}$~\Msun. The normalized impact parameter (\nrhocl) of our quasar-cluster pairs ranges from $1.0 - 5.0$ by design, with a median of $3.6$. Finally, the median impact parameter of our sample is 2.4 Mpc.

\subsection{Continuum normalization} 
\label{subsec:conti}

We found that the default SDSS principal component analysis (PCA) quasar continua are not adequate for the analysis presented here. We, therefore, improve upon the default SDSS continua of the quasars, following the steps described below: 

(a) First, we normalize each spectrum by the corresponding PCA continuum. We considered the wavelength range in the observed frame between 3600 to 5300~\AA\ only. This range is sufficient to cover the \MgII\ doublet along with the \FeII~$\lambda 2600$ and \MgI~$\lambda 2852$ transitions from all of our clusters. 

\smallskip 
(b) Next, we performed an iterative boxed sigma-clipping with asymmetric sigma-levels on the PCA-continuum-normalized spectrum. This ensures efficient clipping of absorption features while retaining the residual emission line features. The number of boxes was chosen based on the presence/absence of strong emission lines within the wavelength range of our interest. The asymmetric sigma-levels were determined based on the median SNR of each spectrum. 

\smallskip 
(c) The resulting clipped spectrum was then smoothed with a median-filter of size 51 pixels to minimize the residual absorption features. The smoothed spectrum was interpolated linearly over the clipped spectral regions.

\smallskip 
(d) We finally fit spline over this interpolated spectrum. The knots of the spline was optimized based on the presence of emission lines and the SNR of each spectrum to minimize over-fitting in the noisy spectral regions. 
\smallskip 

In Fig.~\ref{fig:conti1}, we present a step-by-step illustration of our continuum-fitting algorithm for the quasar SDSS J013346.72$-$002534. In Fig.~\ref{fig:conti2}, we show three examples of quasar spectra with SDSS default continua along with the improved continua from our new algorithm. It is apparent from Figs.~\ref{fig:conti1} and \ref{fig:conti2} that our new adopted continua are significantly improved compared to the default SDSS continua.

\section{Results} 
\label{sec:result}

In this study, we use the spectral stacking technique \citep[see e.g.,][]{Muzahid2021} to probe the CCM of the clusters in our sample. For each quasar-cluster pair, we first shift the normalized quasar spectrum to the rest-frame of the cluster by dividing the observed wavelengths by ($1+z_{\rm cl}$). We then calculate the mean and median fluxes in bins of $1$~\AA. To avoid a few bad pixels, we use $50\sigma$ clipping. We confirm that our results are not sensitive to either the bin size or the $50\sigma$ clipping used here. We focus on the rest-frame 2500--3000~\AA\ spectral range, which covers both the \FeII~$\lambda 2600$ and \MgI~$\lambda 2852$ in addition to the targeted \MgII~$\lambda 2796,2803$ transitions. In the top panel of Fig.~\ref{fig:stackprofiles}, we show the number of quasar-cluster pairs contributing to the stacks at each rest-frame wavelength bin. As expected, all the quasar-cluster pairs in our sample are contributing to the stacked profiles near the \MgII\ wavelengths. However, the number of pairs is reduced for the \FeII~$\lambda~2600$ line. 

%------------------------------------------------------------------- 
\begin{table*}
\begin{adjustwidth}{-2cm}{}
\scriptsize
\caption{Summary of the measurements performed on the mean and median stacks.} 
\label{tab:results} 
\begin{tabular}{@{}cccccccccccc@{}} 
\hline \hline
\multicolumn{1}{c}{Sample} & $N_{\rm pairs}$ & \zcl\ & $M_{500}$ & $R_{500}$ & \rhocl  & \nrhocl & \multicolumn{2}{c}{REW(\MgII) in \AA}  &    \\
& & & ($10^{14}~\mathrm{M}_{\odot}$) & (Mpc) & (Mpc) & & Mean & Median &  ... \\ 
(1) & (2) & (3) & (4) & (5) & (6) & (7) & (8) & (9) & \\
\hline
Full    &  79485  &  0.54(0.46$-$0.62)  &  1.47(1.10$-$2.31)  &  0.65(0.59$-$0.75)  &  2.40(1.46$-$3.16)  &  3.6(2.2$-$4.6)  &  0.034$\pm$0.005  &  0.010$\pm$0.003 &   ... \\ 
Ex-CGM  &  63210  &  0.54(0.47$-$0.63)  &  1.49(1.11$-$2.34)  &  0.65(0.59$-$0.76)  &  2.42(1.48$-$3.17)  &  3.6(2.2$-$4.6)  &  0.031$\pm$0.005  &  0.010$\pm$0.003 &   ... \\ 
\hline 
\end{tabular} 
\end{adjustwidth}
\end{table*}
%-------------------
\begin{table*}
\vspace{-0.24in}
\begin{adjustwidth}{-2cm}{}
\scriptsize
\begin{tabular}{@{}cccccccccc@{}} 
\hline
\multicolumn{2}{c}{$\log_{10}N$(\MgII)/cm$^{-2}$} 
& \multicolumn{2}{c}{REW(\FeII) in \AA} & \multicolumn{2}{c}{$\log_{10}N$(\FeII)/cm$^{-2}$} & \multicolumn{2}{c}{$\sigma_{v}$ (\kms)} &  \multicolumn{2}{c}{$V_{0}$ (\kms)} \\
Mean & Median & Mean & Median & Mean & Median & Mean & Median & Mean & Median  \\ 
(10) & (11) & (12) & (13) & (14) & (15) & (16) & (17) & (18) & (19) \\ 
\hline
11.72$\pm$0.06  &  11.22$\pm$0.14  &  0.010$\pm$0.006  &  0.008$\pm$0.003  &  11.86$\pm$0.24  &  11.75$\pm$0.17  &  355$\pm$36  &  317$\pm$58  &  55 $\pm$116  &  64 $\pm$73  \\
11.68$\pm$0.08  &  11.19$\pm$0.15  &  0.014$\pm$0.006  &  0.008$\pm$0.004  &  11.98$\pm$0.19  &  11.75$\pm$0.19  &  426$\pm$75  &  304$\pm$59  &  146$\pm$149  &  105$\pm$80 \\

\hline
\end{tabular}
\end{adjustwidth}
\begin{tablenotes}\small
\item Notes -- (1) Sample name used in stacking (see Section~\ref{sec:result}). (2) Number of quasar-cluster pairs for a given sample. (3), (4), (5), (6), and (7) median values of cluster redshift, $M_{500}$, $R_{500}$, \rhocl, and \nrhocl~ respectively. The values in the parenthesis for the parameters listed from (3) $-$ (7) indicate the 16 and 84  percentiles of the parameter distribution. (8) \& (9) REWs of \mgii line measured between 2785$-$2815~\AA~ from the mean and median stack spectra, respectively. (10) \& (11) Column density estimated using the \MgII\ REWs in (8) and (9), respectively, assuming lines are in the linear part of the COG. (12) \& (13) REWs of \feii line measured between  2589$-$2611~\AA~ from the mean and median stack profiles, respectively. (14) \& (15) Same as (10) \& (11) but for \feii line. (16) \& (17) Line widths calculated from three-component Gaussian fits to the \feiia and \mgiiab lines for the mean and median stacks, respectively. We tied the line widths of the \mgii and \feii lines. (18) \& (19) Line centroids obtained from the fitting for the mean and median stack spectra, respectively. 
\end{tablenotes}
\end{table*} 
%-------------------------------------------------------------------

The mean and median stacked spectra are shown in the bottom and middle panels of Fig.~\ref{fig:stackprofiles}, respectively. The \MgII\ line is clearly detected in both the mean and median stacked spectra with total REW (\ewmgii) of $0.034\pm0.005$~\AA\ and $0.010\pm0.003$~\AA, indicating $7\sigma$ and $3\sigma$ detection significance for the mean and median stacks. The two members of the \MgII\ doublet are barely resolved in the stacked spectra. The REW values are calculated after re-normalizing the stacked spectra by the corresponding pseudo-continua. The pseudo-continuum levels for the mean and median stack spectra are determined by fitting, respectively, 1st and 2nd order polynomials to the line-free regions. Note that the pseudo-continua are consistent with the overall flux decrements when we randomized the cluster redshifts. The uncertainties in the REW are quadrature addition of the pseudo-continuum placement uncertainty and the statistical uncertainty calculated from 200 bootstrap realizations of the full sample.
 
Assuming that the observed \MgII\ line falls on the linear part of the curve-of-growth (COG), and only $2/3$ of the total REW is owing to the \MgII~$\lambda2796$ line \footnote{Since the \MgII~$\lambda2796$ transition is 2 times stronger compared to the \MgII~$\lambda2803$ transition.}, we obtained \ewmgiia of $0.023\pm 0.003$~\AA\ ($0.007\pm 0.002$~\AA) and column density of $\log_{10} N(\MgII)/\rm cm^{-2} = 11.72\pm 0.06$ ($11.22\pm 0.14$) for the mean-stacked (median-stacked) spectrum. 

Besides \MgII, marginal \FeII~$\lambda 2600$ absorption line is detected with mean (median) REW of $0.010\pm 0.006$~\AA\ ($0.008\pm 0.003$~\AA). The corresponding column density is $\log N(\FeII)/\rm cm^{-2} = 11.86\pm 0.24$ ($11.75\pm 0.17$) for the mean (median) stack. We fit the \FeII\ and \MgII\ lines simultaneously with a three-component Gaussian centered at the \feiia and \mgiiab lines with the centroids separated by the rest-frame wavelengths of the transitions and by keeping the widths ($\sigma_v$) tied and amplitudes free. The best-fit Gaussian has a velocity centroid ($V_0$) of $55\pm 116$~\kms\ for the mean-stacked profile ($64\pm 73$~\kms\ for the median), which is fully consistent with $0$~\kms. The measured line widths are $355 \pm 36$~\kms\ and $317 \pm 58$~\kms\ for the mean and median stacks, respectively. The measurements performed on the stacked profiles are summarized in Table~\ref{tab:results}. Finally, we do not detect any \MgI\ absorption. Given the extremely high SNR of the stacked spectrum, the non-detection of \MgI\ implies mean REW(\MgI) $< 0.003$~\AA\ ($3\sigma$), assuming that the undetected line is spread over 8 pixels (2.355$\times\sigma_{v} \approx 840$~\kms). \par

Next, we produced stacks for three mass and three redshifts bins with a nearly equal number of quasar-cluster pairs in each bin. We did not find any significant dependence of \MgII\ REW on either mass or redshift, likely due to the narrow ranges of mass and redshift probed here (see 16 and 84 percentiles of the sample distribution from Table~\ref{tab:results}). Similarly we split the full sample in three bins of \rhocl\ and \nrhocl\ to construct the \MgII\ REW-profile. The details of the bins and the measurements are summarized in supplementary Table~S1.\footnote{The mean and median stacked profiles for each bin are also presented in the supplementary material.} The REW-profiles for the mean/median \MgII\ absorption are shown in Fig.~\ref{fig:result_bins}. A decrease in the \MgII\ REW with increasing \rhocl\ and \nrhocl\ is evident from the figure.\par

\section{Discussion and Conclusion} 
\label{sec:discussion}

\begin{figure*}
\begin{center}
     \includegraphics[width=0.95\textwidth]{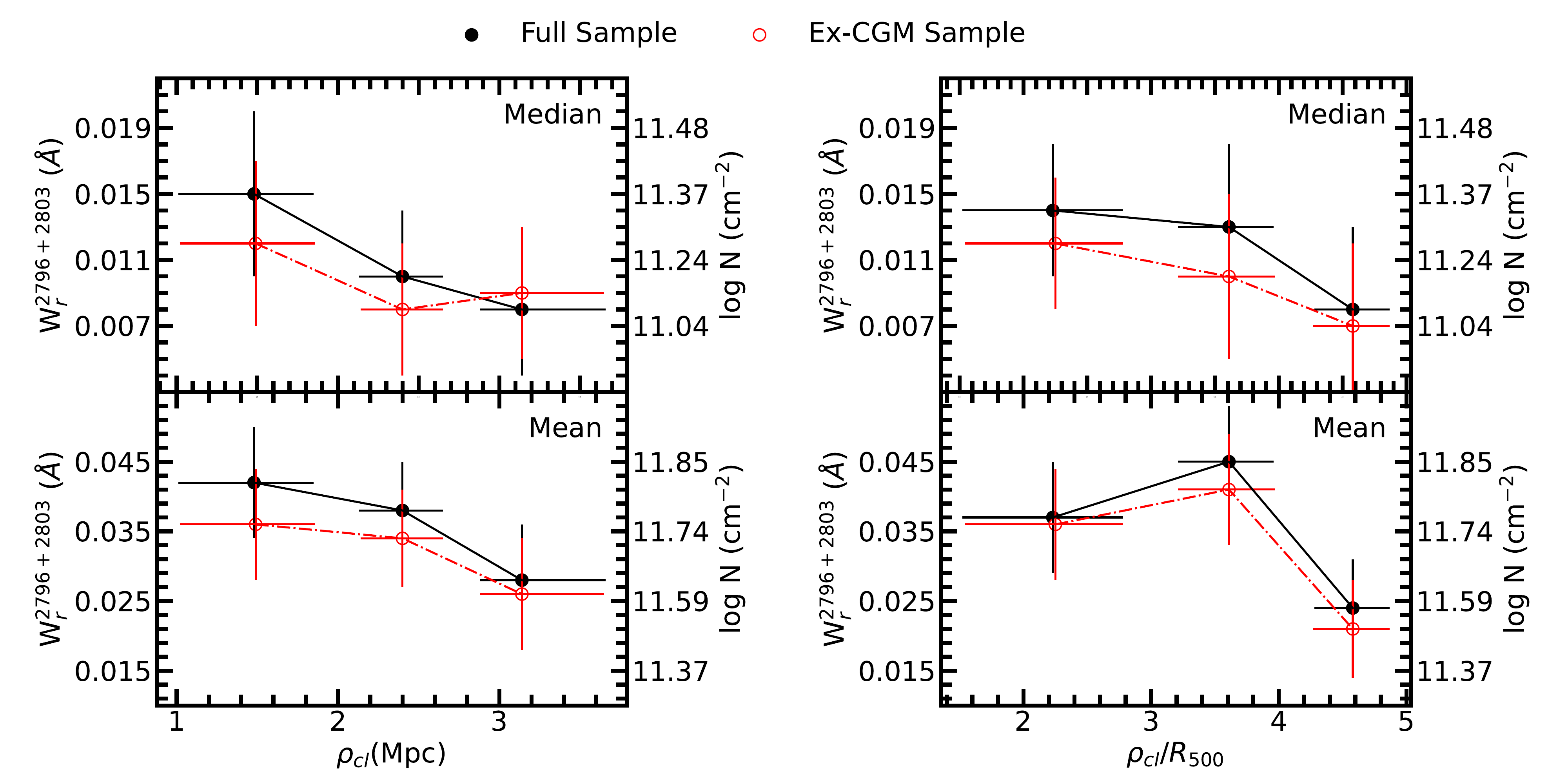} 
     \caption{ \emph{Left:} REW of \MgII\ obtained from the mean (bottom) and median (top) stacked spectra versus impact parameter (\rhocl). The REWs measured for the full sample are shown in black solid circles. The measurements for the ``Ex-CGM'' sub-sample, i.e., after excluding the quasar-cluster pairs that can have significant contribution from the CGM of galaxies in the cluster outskirts near the quasar sightlines (see Section~\ref{sec:discussion}), are shown in open red circles. The error bars along the y-axis indicate the $1\sigma$ range obtained from combining the statistical uncertainty from 200 bootstrap realizations and pseudo-continuum placement error}, while the error bars along the x-axis represent the 68 percentile range for \rhocl\ bin. The right y-axes show the \MgII\ column density corresponding to the REW on the left for the linear part of the COG. \emph{Right:} Same as the left panel but for \nrhocl.
     \label{fig:result_bins}
\end{center}
\end{figure*}

We detect $7 \sigma$ significant \MgII\ and marginal \FeII\ absorption in the mean stacks  arising from the CCM of $z\approx0.5$ clusters with a median mass of $\approx10^{14.2}$~\Msun. Our results show that cluster outskirts (1--5~$R_{500}$) are enriched with heavy elements. The mean and median \MgII\ REW measured from our stacks are consistent with the values inferred by \citet[see their Eqn.~5]{anand2022cool} at the median \nrhocl ($= 3.6$)  within $2\sigma$.

The prevalence of metal-rich, cool/warm-hot gas in the outskirts is also seen in theoretical studies of Virgo-like and Coma-like clusters \citep[see][]{Emerick2015,Butsky2019}. These authors suggested that the cool gas in the outskirts are associated with the infall of pristine gas from the IGM and/or with early-stage stripping of the CGM of infalling galaxies. As a consequence, the gas in the outskirts shows a wide range in metallicity \citep[see Fig.~5  of][]{Butsky2019}. The existence of metal-rich gas in the cluster outskirts, as we observed here, can also be explained by a uniform early metal enrichment scenario where deposition and mixing of heavy elements occurred well before the cluster formation when the star formation and AGN activities were on their peaks \citep[i.e., $z \gtrsim 2$; see][]{Simionescu2015,Urban2017}. 
Alternatively, theoretical models of clusters predict that cool gas ($\sim$ 10$^{4}$ K) can condense out of the hot ICM plasma ($\sim10^{7}$~K)  due to thermal instability when the cooling time scale ($t_{\rm cool}$) reaches 10 times below the gravitational free-fall timescale \citep[$t_{\rm ff}$, e.g.,][]{McCourt2012,Sharma2012,Voit2017}. However, such a process is not important at distances $\gtrsim 100$~kpc from cluster centres owing to the large $t_{\rm cool}/t_{\rm ff}$ ratio  \citep[$\gg 10$, e.g.,][]{Hogan2017} due to the rapidly decreasing density of the hot gas with distance. We recall that the regions we probed here are even further away from cluster centres.   

We find a decreasing trend between \MgII\ REW and \rhocl\ and \nrhocl, likely due to the lower density, metallicity, and covering fraction of cool gas at large radii \citep[]{Emerick2015,Butsky2019}. A similar declining trend is well-known in the CGM literature for decades \citep[e.g.,][]{Lanzetta1990,Chen2010,Nielsen2018,Dutta2021}. Adopting the relation presented in \citet{Dutta2021}, we obtained galactocentric distances of $200-300$~kpc for the \MgII\ REW we measured from the stacked profiles, provided that the absorption signal is dominated by the CGM of galaxies in the outskirts. To quantify the possible contribution from the CGM of galaxies in the outskirts, we searched for galaxies in the SDSS using $CasJobs$\footnote{\color{blue}{https://skyserver.sdss.org/casjobs/}}. We limited our search to the galaxies with ``photoErrorClass'' flags of $-1$, $1$, $2$, and $3$ with reliable photometric measurements. These galaxies have typical uncertainty of $\approx 0.03$ in the photometric redshifts \footnote{ This corresponds to a velocity of $\approx6000$~\kms\ at the median cluster redshift of 0.5.} \citep[]{Beck2016}. We excluded $16,275$ pairs from our analysis for which at least one galaxy is detected within 300~kpc radius centred on the background quasar with a photometric or spectroscopic redshift consistent with the cluster redshift within $\pm6000$~\kms.\footnote{ We have identified nearly $20,000$ ($\sim L_{*}$) galaxies satisfying these conditions with median $r$-band apparent and absolute magnitudes of $\approx21.6$ and $\approx -21.4$, respectively.} The measurements obtained for this sub-sample  after excluding the possible CGM contribution (``Ex-CGM'') are summarized in Table~\ref{tab:results}.\footnote{The mean and median composite spectra for this sample are shown in supplementary Fig.~S1.} The mean/median \MgII\ (and \FeII) REWs for the ``Ex-CGM'' sub-sample are consistent within 1$\sigma$ with the full sample. Moreover, the ``Ex-CGM'' sub-sample shows trends similar to the full sample (see Fig.\ref{fig:result_bins}). We, thus, conclude that the absorption is not dominated by the CGM of bright ($\sim L_*$) galaxies in the outskirts near the quasar sightlines. However, we cannot rule out the possibility of contribution from the CGM of low-mass/faint galaxies. In passing, we note that the stacked spectra for the $16,275$ quasar-cluster pairs with possible CGM contributions exhibit a stronger \mgii\ absorption with mean and median REWs of 0.053$\pm$0.007~\AA\ and 0.020$\pm$0.003~\AA, respectively.\par

We run grids of photoionization models using {\sc cloudy} \citep[v17.02;][]{Ferland2017} with metallicity ($\log_{10} Z$), neutral hydrogen column density ($\log_{10} N(\HI)/$\sqcm), and density ($\log_{10} n_{\rm H}/\rm cm^{-3}$) ranging from $-2.0$~--~$0.0$, $14.0$~--~$18.0$, and $-3$~--~$0.0$, respectively, in steps of 0.5 dex. We assumed that the absorbing medium has   solar relative abundance, and is subjected to the extragalactic UV background radiation at $z= 0.5$ obtained from \citet{Haardt2012}. These models suggest a moderate density of $n_{\rm H} > 10^{-2.0}~\rm cm^{-3}$ to reproduce the observed mean $N(\MgII)/N(\FeII)$ ratio, for the whole range of $N(\HI)$ and metallicity explored. Moreover, the metallicity required to match the observed mean \MgII\ and \FeII\ column densities is $\log_{10} Z \gtrsim -1.5$, provided $\log_{10} N(\HI)/{\rm cm^{-2}} < 17.2$ (sub-LLS) and $\log_{10} n_{\rm H}/{\rm cm^{-3}} > -2.0$. The density and metallicity constraints obtained from these simple ionization models favor an origin similar to circumgalactic gas as opposed to pristine filamentary accretion for the absorption signal detected here. However, the strength of the \MgII\ absorption for the Ex-CGM sample is consistent with the full sample, implying that the signal is not dominated by the gas locked in the CGM of bright galaxies near the background quasars. These two facts suggest that the \MgII\ absorption likely to arise from stripped gas, and that stripping can be important at such large distances (\rhocl~$\approx 2.4$~Mpc). The simulations of \citet{Butsky2019} also showed that CGM stripping is relevant for galaxies as far as $4R_{200}$. Similarly, based on the little correlation between the \mgii absorbers and properties of their nearest galaxies within $R_{200}$ of clusters and the fact that the mean absorber$-$nearest galaxy separation is larger ($>200$ kpc) than the typical size of star-forming disk of the galaxy, \citet{anand2022cool} argued that the cool gas in the outskirts can in part arise from stripping.
\par

Using the empirical relationship between REW(\MgII) and $N(\HI)$ obtained by \citet{Lan2017}, we obtained $N(\HI) \approx 10^{16.8}$~$\rm cm^{-2}$ for the mean REW(\MgII) of $0.034$~\AA, suggesting a sub-LLS environment for the absorbing gas. Such high \HI\ column densities are consistent with the observations of \citet{Muzahid2017} but rarely observed in the outskirts of Virgo/Coma clusters \citep[see Fig. 3 of][]{Muzahid2017} and in hydrodynamical simulations \citep[see e.g.,][]{Emerick2015}. It may be possible that the empirical relation is not valid for such low \MgII\ REW. Dedicated UV absorption line programs using $HST$ are essential to map the distribution of neutral gas and to probe the multiphase structure of cluster outskirts.

In conclusion, using a large sample of quasar-cluster pairs, we report strong \MgII\ and marginal \feii\ absorption in the composite spectra of quasars arising from the CCM of $z\approx0.5$ clusters with a median mass of $\approx10^{14.2}$~\Msun. This is the first statistical detection of metal-enriched, cool gas in the outskirts of galaxy clusters. We discussed the possibilities, such as filamentary accretion from the IGM, uniform early metal enrichment, ICM cooling, stripped of galactic materials, and gas locked in the CGM, that could give rise to the detected absorption. We showed that the CGM of bright galaxies ($\sim L_{*}$) in the outskirts does not dominate the signal. 
The chemical and physical conditions inferred from simple photoionization models, however, suggest that the absorption is arising from gas with moderate density and metallicity which are typical of the CGM. We therefore speculate that the absorption stems from stripped of galactic materials and that stripping may be important at large clustocentric distances ($\approx 3-4 R_{500})$ as seen in some recent cosmological hydrodynamical simulations. But the possibility of contribution from the CGM of low-mass/faint galaxies to the detected absorption signal cannot be ruled out. Future large galaxy surveys with deeper photometric and/or spectroscopic completeness can shed light on this. We plan to determine the covering fraction of \MgII\ in the cluster outskirts and investigate environmental effects on the CGM using the $\approx$20,000 galaxies in the outskirts identified within 300~kpc of the quasar sightlines in the future.

\vskip0.4cm 
We thank Raghunathan Srianand and Aseem Paranjape for useful discussion. We acknowledge the use of High performance computing facility PEGASUS at IUCAA. Funding for SDSS-III has been provided by the Alfred P. Sloan Foundation, the Participating Institutions, the National Science Foundation, and the U.S. Department of Energy Office of Science. The SDSS-III web site is http://www.sdss3.org/. SDSS-III is managed by the Astrophysical Research Consortium for the Participating Institutions of the SDSS-III Collaboration including the University of Arizona, the Brazilian Participation Group, Brookhaven National Laboratory, Carnegie Mellon University, University of Florida, the French Participation Group, the German Participation Group, Harvard University, the Instituto de Astrofisica de Canarias, the Michigan State/Notre Dame/JINA Participation Group, Johns Hopkins University, Lawrence Berkeley National Laboratory, Max Planck Institute for Astrophysics, Max Planck Institute for Extraterrestrial Physics, New Mexico State University, New York University, Ohio State University, Pennsylvania State University, University of Portsmouth, Princeton University, the Spanish Participation Group, University of Tokyo, University of Utah, Vanderbilt University, University of Virginia, University of Washington, and Yale University.

\bibliography{references}{}
\bibliographystyle{aasjournal}

\appendix

\setcounter{section}{0}
\setcounter{table}{0}
\setcounter{figure}{0}

\counterwithin{figure}{section}
\renewcommand{\thefigure}{A\arabic{figure}}
\renewcommand{\thetable}{A\arabic{table}}  
\renewcommand{\thesection}{A\arabic{section}}

\begin{figure*}
            \includegraphics[width=0.93\textwidth]{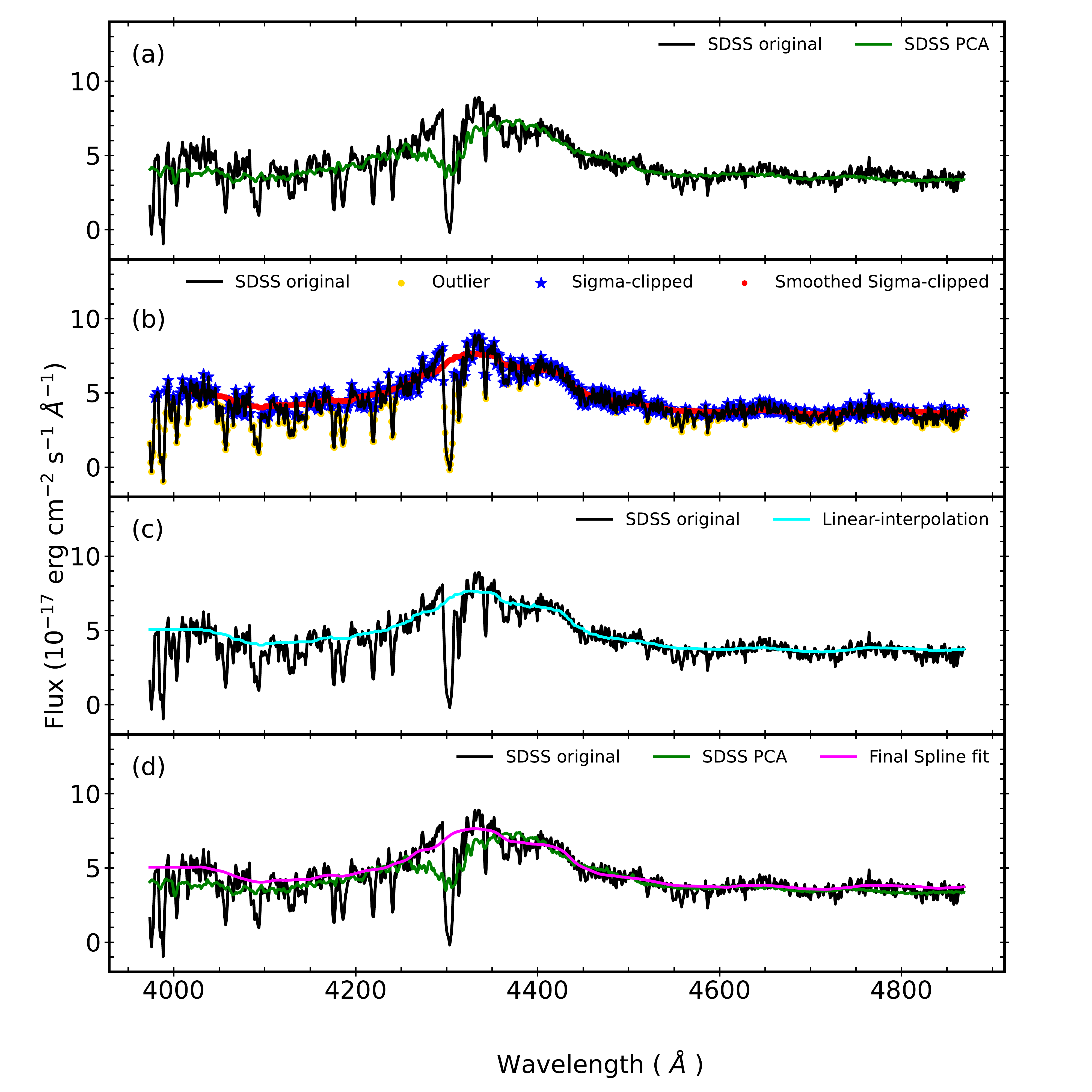}
     \caption{An illustration of our continuum-fitting algorithm. First, we normalize the original spectrum (black, \emph{a}) with the default SDSS PCA continuum (green, \emph{a}). After that, we perform an iterative boxed-sigma clipping with  asymmetric levels on the upper and lower sides of the normalized continuum. The yellow data points, marked as outliers, in the panel (\emph{b}) indicate the location of possible absorption features that are clipped by our clipping algorithm. The blue scatter points in the panel (\emph{b}) show the sigma-clipped spectrum.  Clipping is performed at 50 and 2 upper and lower sigma levels, respectively, after dividing the spectrum in 10 boxes of 90~\AA\ width. We smooth these sigma-clipped data points using 51 pixels to minimize residual absorption (red, \emph{b}). We then linearly interpolate the red data points in (\emph{b}) to generate a smooth model (cyan, \emph{c}). Finally, we fit a spline (of 70 knots in this case) over this interpolated model resulting in the final continuum fit (magenta, \emph{d}). Note that our adopted continuum in magenta is significantly improved compared to the default SDSS continuum shown in green, particularly near the strong absorption at $\approx4300$~\AA.} 
     \label{fig:conti1} 
\end{figure*}
\newpage

\begin{figure*}
            \includegraphics[width=1.0\textwidth]{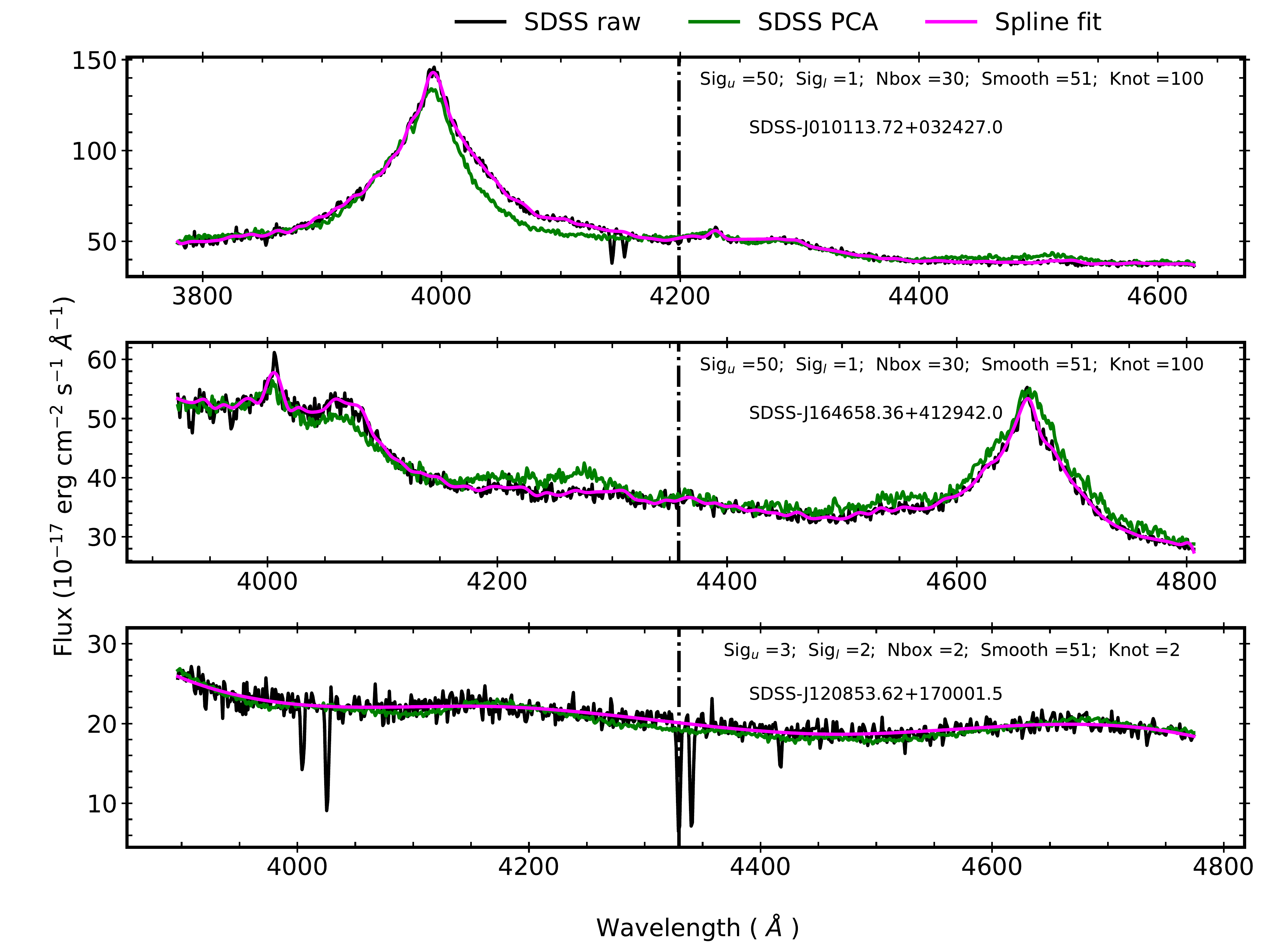}
     \caption{Examples of SDSS spectra (in black), default SDSS continua (in green) and our improved continua (in magenta). The spectral regions correspond to $\pm30,000$~\kms\ around the cluster redshifts. The upper and lower sigma levels ($\rm Sig_u/l$), the number of boxes ($\rm Nbox$) within which each spectrum is divided, the number of pixels used for smoothing ($\rm Smooth$) and the number of knots ($\rm Knot$) used for the spline fitting are given on the top right corner of each panel. At the centre of each panel, the dashed line corresponds to the cluster's \MgII\ doublet location. Our new algorithm significantly improved our adopted continuum in all three cases compared to the SDSS PCA continuum.} 
     \label{fig:conti2} 
\end{figure*}

\setcounter{section}{0}
\setcounter{table}{0}
\setcounter{figure}{0}

\renewcommand{\thefigure}{S\arabic{figure}}
\renewcommand{\thetable}{S\arabic{table}}  
\renewcommand{\thesection}{S\arabic{section}}

\begin{center}
{\Large \sc Supplementary Material (online-only)}\\
\end{center}

\section{Stacks for different sub-samples}

\begin{figure*}
      \includegraphics[width=1.0\textwidth]{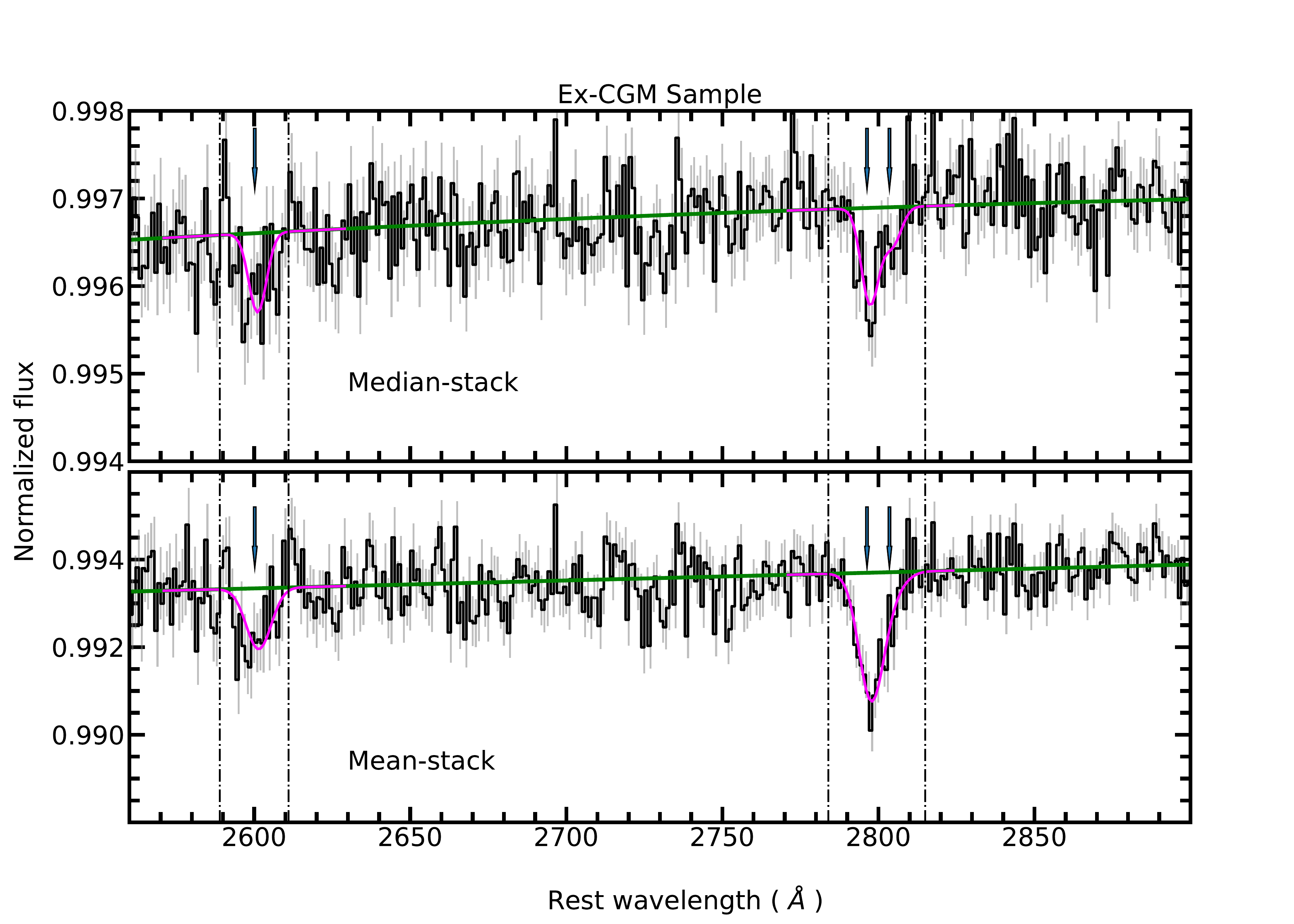}
     \caption{\emph{Top:} Median composite spectrum of the background quasars at the rest-frame of the  clusters for the ``Ex-CGM" sample (excluding the $16,275$ quasar-cluster pairs where the quasars may be  probing the CGM of galaxies in the outskirts). The $1\sigma$ errorbars on flux are estimated from 200 bootstrap samples and are shown in grey. The solid green curve shows the best-fit pseudo continuum. The best-fit \mgii and \feii absorption profiles are shown in magenta curves. The line centroids are shown by the arrows. The four dashed horizontal lines two each around \feii and \mgii corresponds to $[2589, 2611]$~\AA\ and $[2784, 2815]$~\AA\ regions used for rest-frame equivalent width measurements. \emph{Bottom:} same as the top panel but for the mean composite spectrum.}
     \label{fig:stackprofilesExcgm} 
\end{figure*}

In Fig.~\ref{fig:stackprofilesExcgm}, we show the mean and median stack spectra of the ``Ex-CGM'' sample excluding the $16,275$ quasar$-$cluster pairs for which we found at least one galaxy within a radius of 300~kpc around the background quasars and having a photometric or spectroscopic redshift consistent with the cluster redshift within $\pm6000$~\kms.\\

In Fig.~\ref{fig:subsample1}, we present the mean (\emph{bottom}) and median (\emph{top}) stacked spectra for three bins of cluster redshift (\zcl) ensuring each bin contains almost an equal number of quasar$-$cluster pairs. The details of theses bins are given in Table~\ref{tab:results_bins}.\\

In Fig.~\ref{fig:subsample2}, we present the mean (\emph{bottom}) and median (\emph{top}) stack spectra for three bins of $M_{500}$ ensuring each bin contains almost an equal number of quasar$-$cluster pairs. The details of theses bins are given in Table~\ref{tab:results_bins}.\\

In Fig.~\ref{fig:subsample3}, we present the mean (\emph{bottom}) and median (\emph{top}) stack spectra for three bins of \rhocl\ ensuring each bin contains almost an equal number of quasar$-$cluster pairs. The details of theses bins are given in Table~\ref{tab:results_bins}.\\

In Fig.~\ref{fig:subsample4}, we present the mean (\emph{bottom}) and median (\emph{top}) stack spectra for three bins of \nrhocl\ ensuring each bin contains almost an equal number of quasar$-$cluster pairs. The details of theses bins are given in Table~\ref{tab:results_bins}.\\

In Table~\ref{tab:results_bins}, we summarize the bin details and their corresponding stacked results for our various sub-samples that are binned in \zcl, $M_{500}$, \rhocl, and \nrhocl.

%%%%%============
\begin{figure*}
      \includegraphics[width=18cm, height=11.5cm]{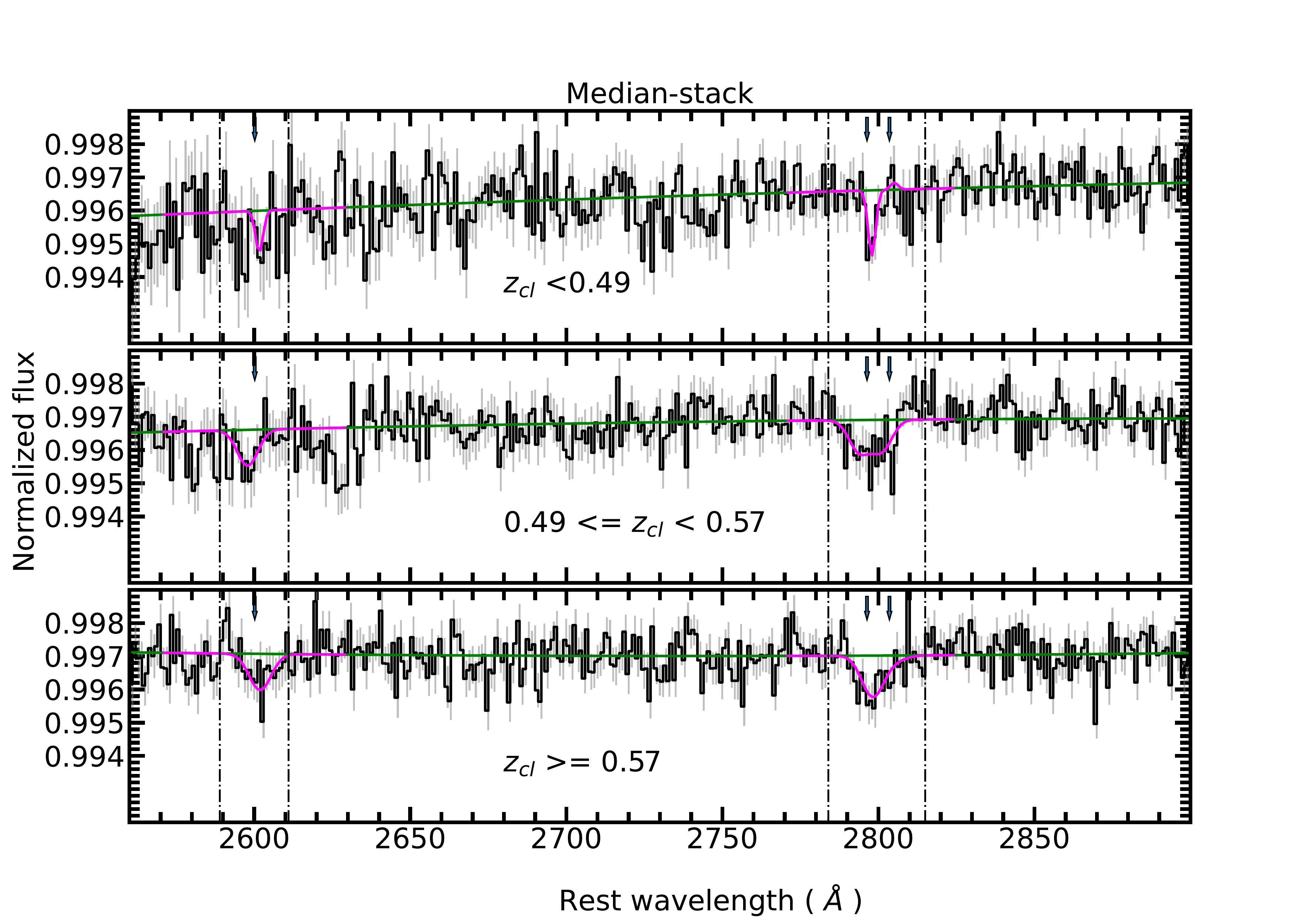}
\end{figure*}
\begin{figure*}
\vspace{-1.2cm}
      \includegraphics[width=18cm, height=11.5cm]{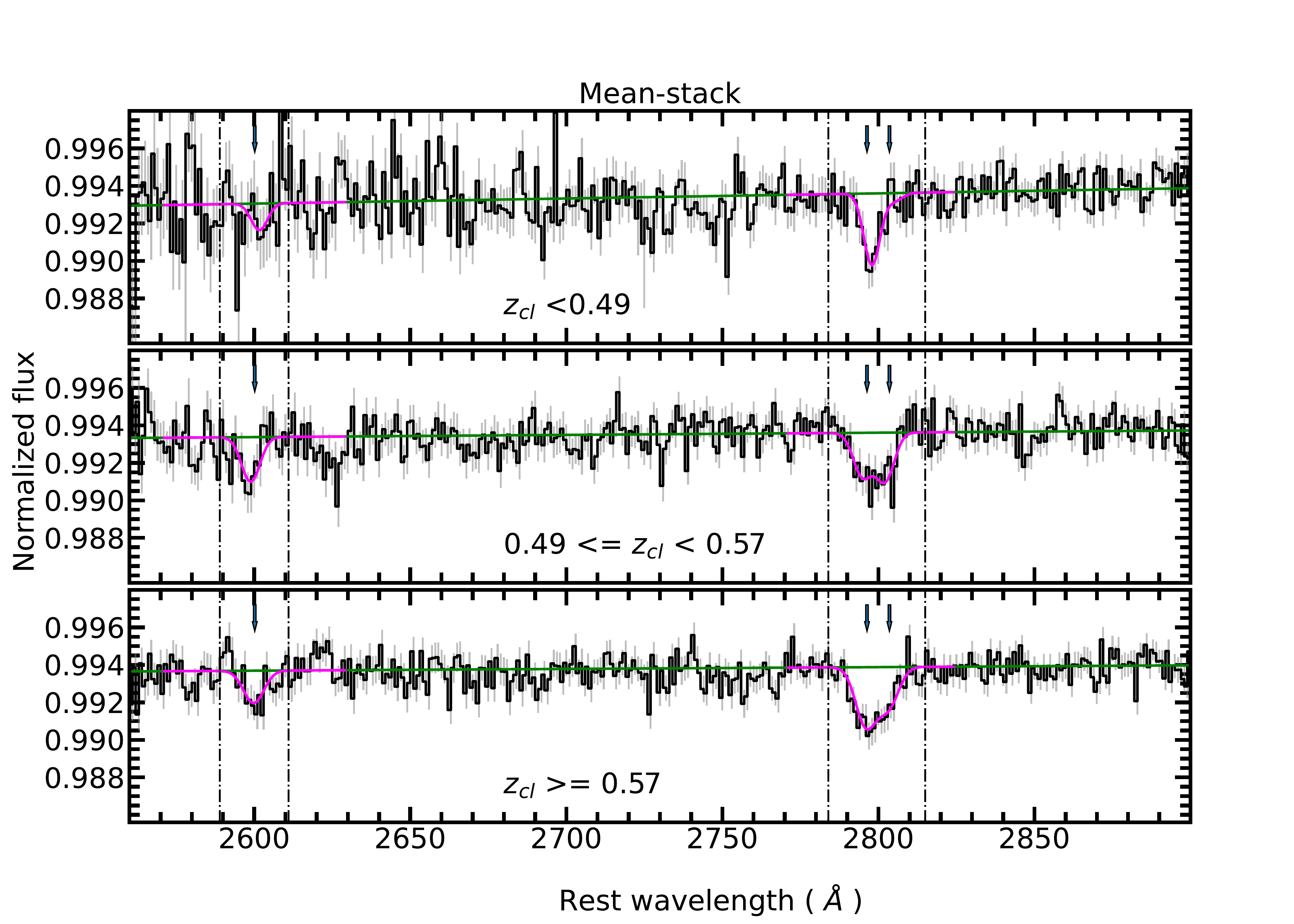}
       \caption{Same as Fig.~\ref{fig:stackprofilesExcgm} but for three different bins of cluster redshift.}
       \label{fig:subsample1} 
\end{figure*}
%%%%%%%%%=========
\begin{figure*}
      \includegraphics[width=18cm, height=11.5cm]{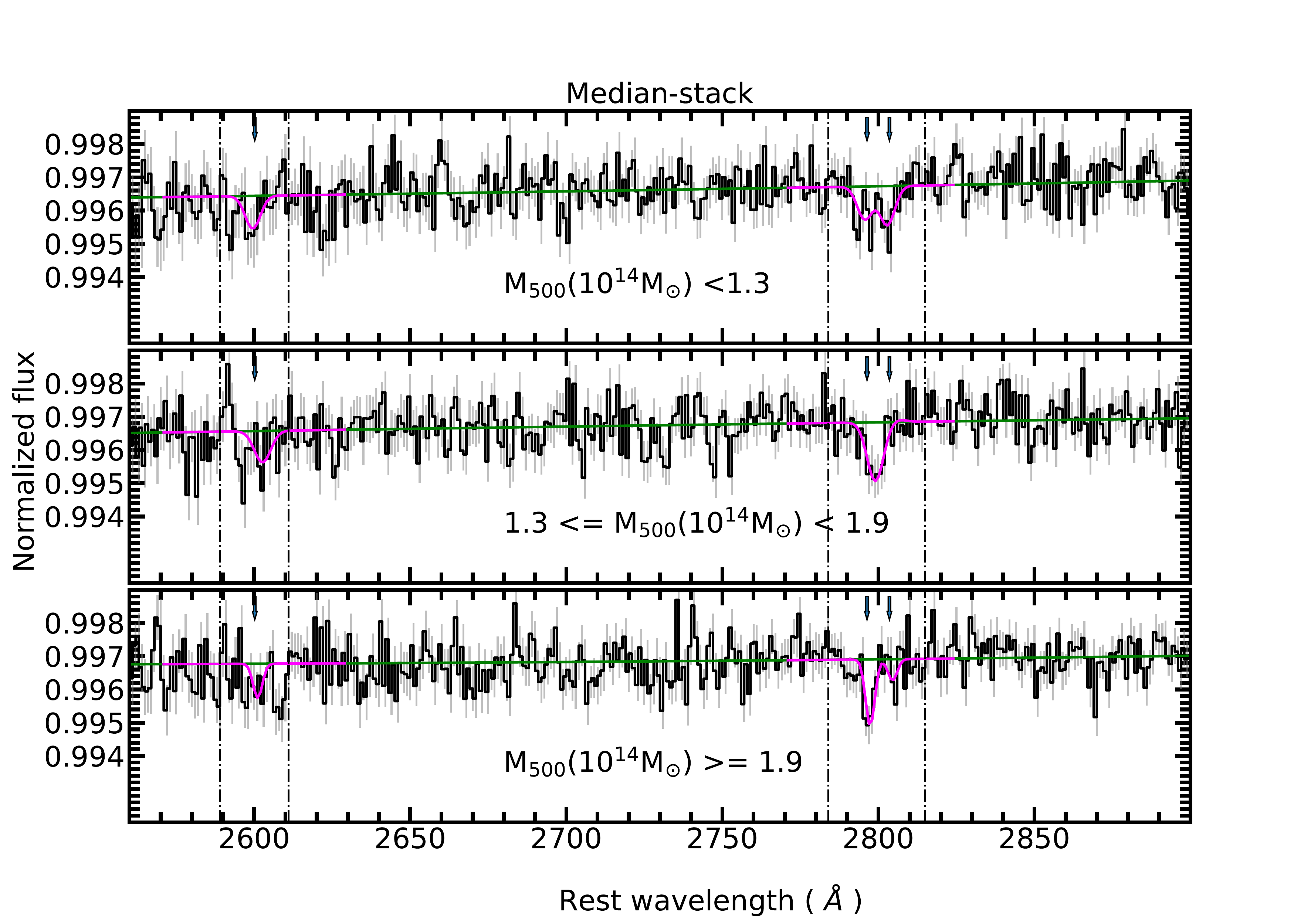}
\end{figure*}
\begin{figure*}
\vspace{-1.2cm}
      \includegraphics[width=18cm, height=11.5cm]{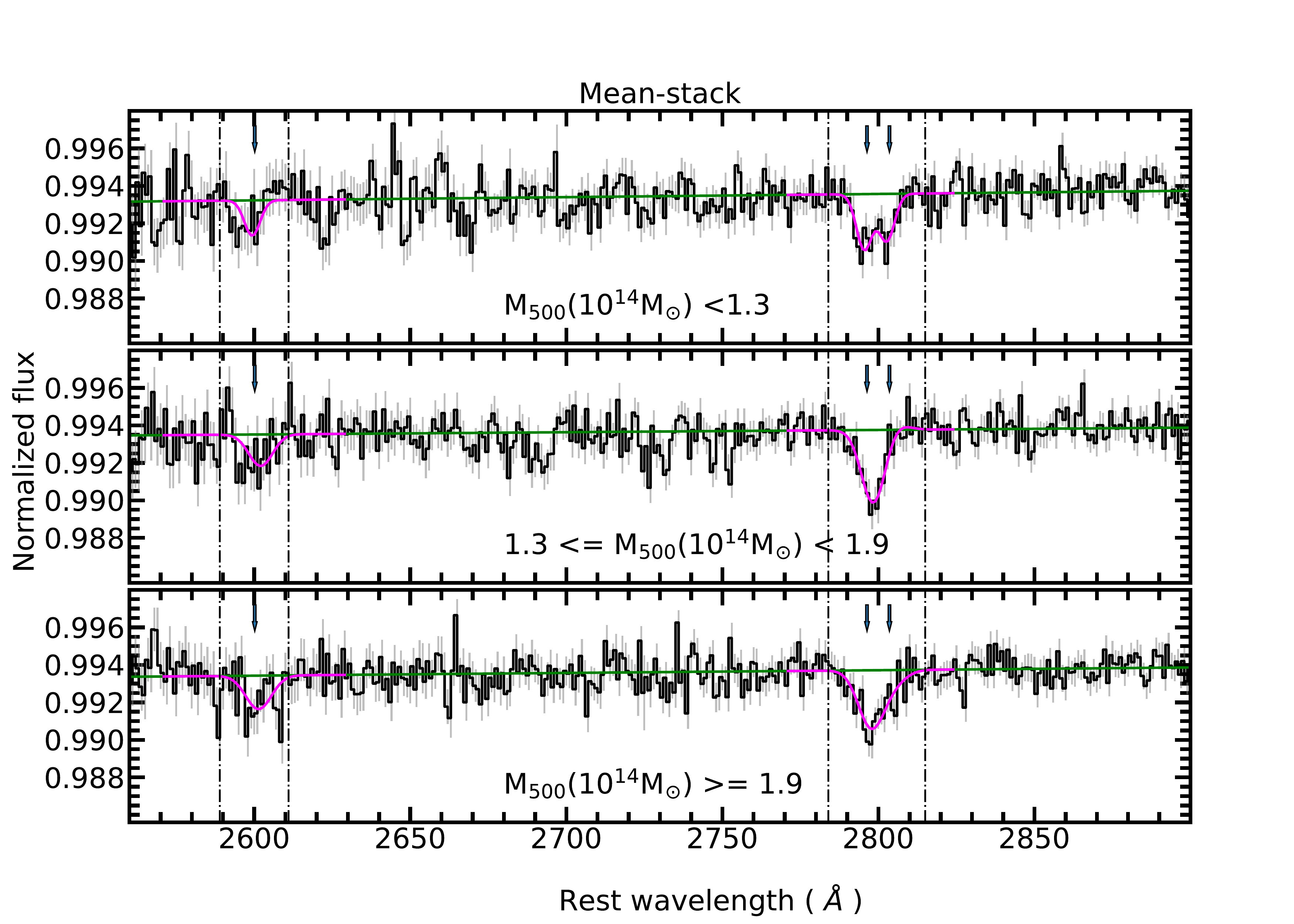}
       \caption{Same as Fig.~\ref{fig:stackprofilesExcgm} but for three different bins of $M_{500}$.}
       \label{fig:subsample2} 
\end{figure*}
%%%%%============
\begin{figure*}
      \includegraphics[width=18cm, height=11.5cm]{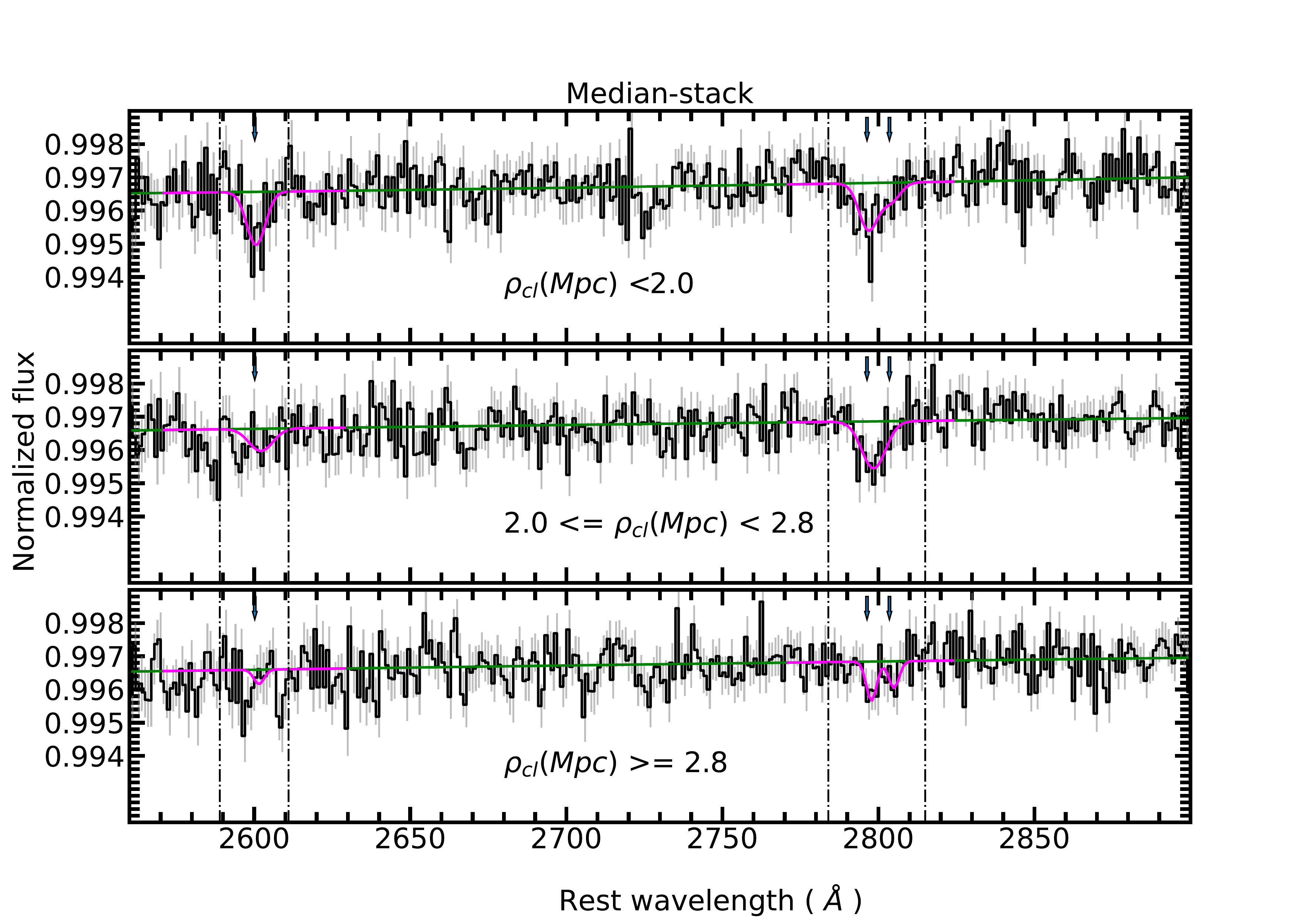}
\end{figure*}
\begin{figure*}
\vspace{-1.2cm}
      \includegraphics[width=18cm, height=11.5cm]{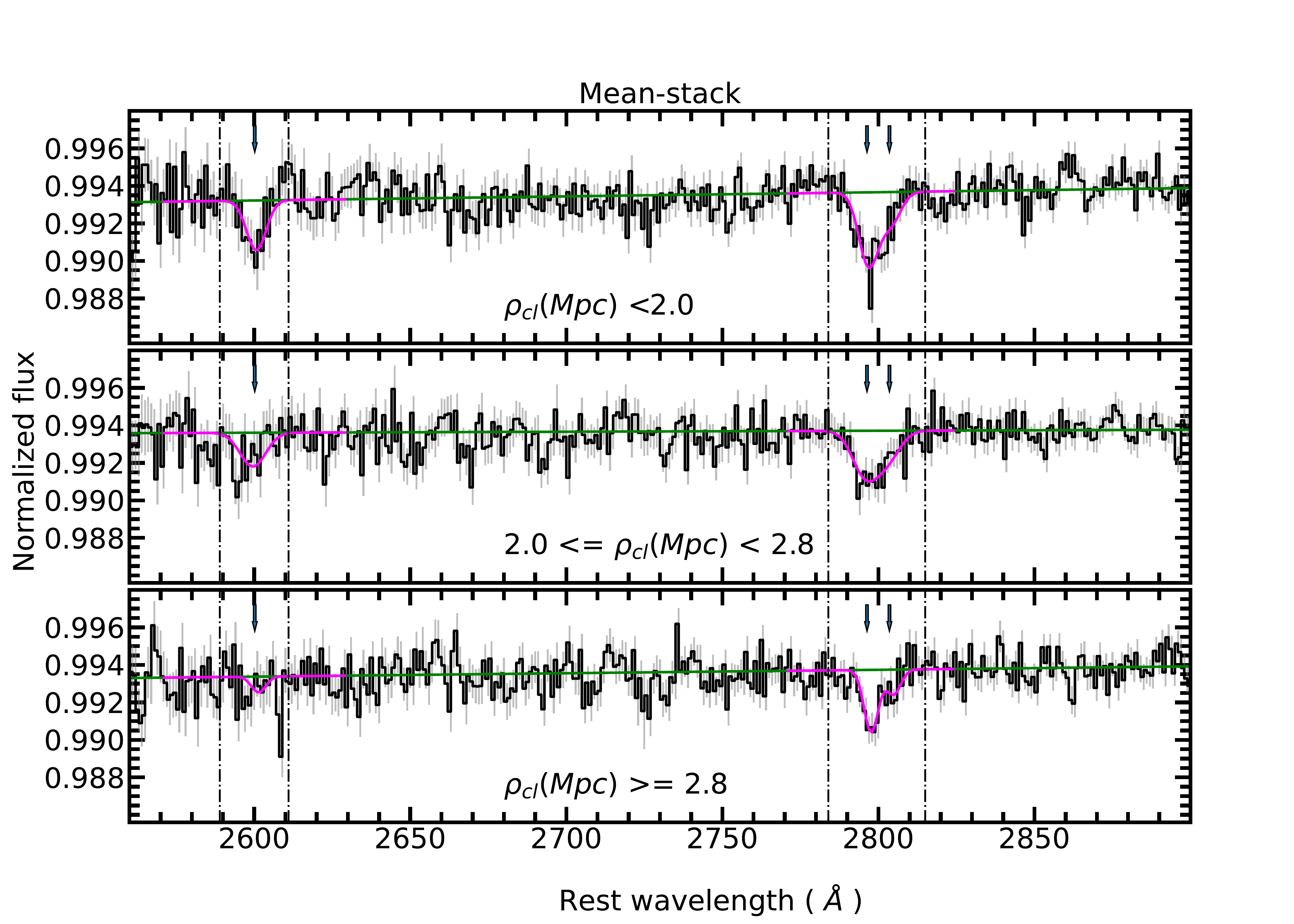}
       \caption{Same as Fig.~\ref{fig:stackprofilesExcgm} but for three different bins of \rhocl.}
       \label{fig:subsample3} 
\end{figure*}
%%%%%============
\begin{figure*}
      \includegraphics[width=18cm, height=11.5cm]{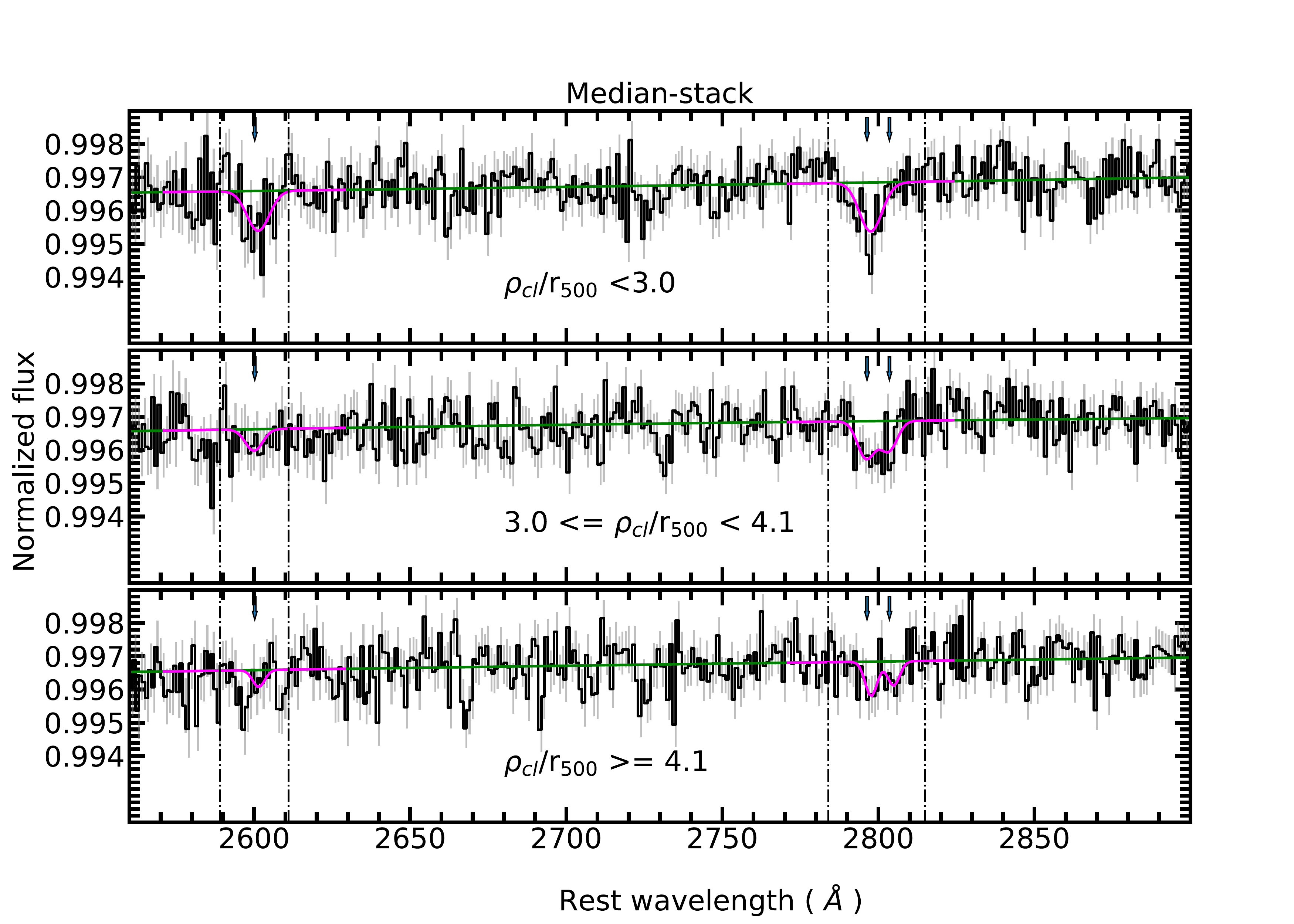}
\end{figure*}
\begin{figure*}
\vspace{-1.2cm}
      \includegraphics[width=18cm, height=11.5cm]{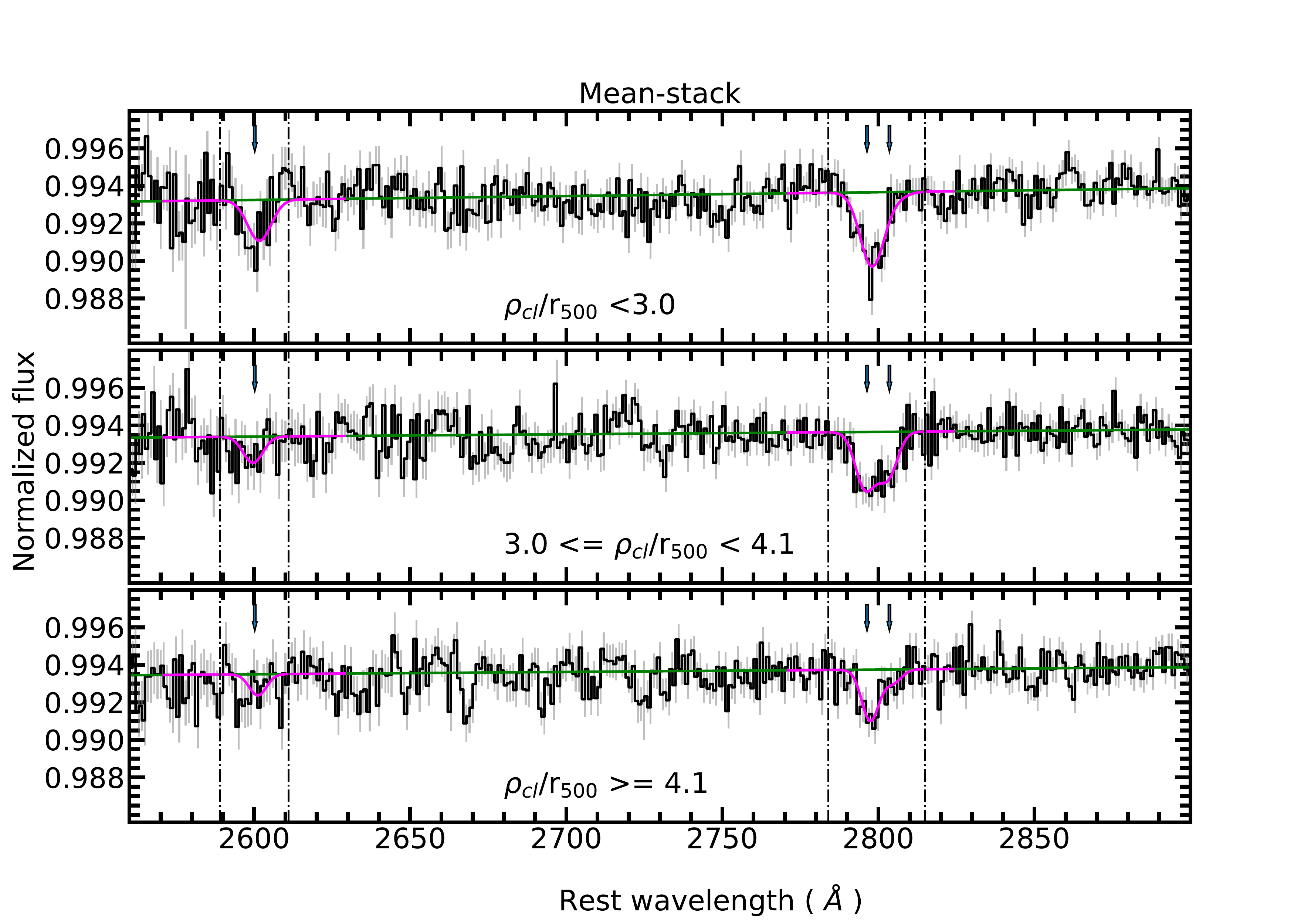}
       \caption{Same as Fig.~\ref{fig:stackprofilesExcgm} but for three different bins of \nrhocl.}
       \label{fig:subsample4} 
\end{figure*}
%%%%%============

\begin{sidewaystable}
\scriptsize
  \caption{Summary of the measurements for the subsamples.}
  \label{tab:results_bins}
\begin{center}
\scalebox{0.8}
{
\begin{threeparttable}
   \begin{tabular}{@{}ccccccccccccccccccc@{}} 
\hline \hline

\multicolumn{1}{c}{Bin} & $N_{\rm pairs}$ & \zcl\ & \multicolumn{2}{c}{REW(\MgII) in \AA}  & \multicolumn{2}{c}{$\log_{10}N$(\MgII)/cm$^{-2}$}  & \multicolumn{2}{c}{REW(\FeII) in \AA} & \multicolumn{2}{c}{$\log_{10}N$(\FeII)/cm$^{-2}$} & \multicolumn{2}{c}{$\sigma_{v}$ (\kms)} &  \multicolumn{2}{c}{$V_{0}$ (\kms)} \\
               &         &                &  Mean & Median & Mean & Median & Mean & Median & Mean & Median & Mean & Median & Mean & Median  \\
 (1)           &  (2)    &  (3)          &   (4)        &    (5)           &   (6)         &   (7)    &  (8)   & (9) & (10)   & (11) & (12) & (13) & (14) & (15)      \\ \hline \hline

$z_{cl}$$<$0.49                                 & 26446 & 0.46$_{0.44}^{0.48}$ & 0.032$\pm$0.009 & 0.012$\pm$0.006 & 11.71$\pm$0.41 & 11.29$\pm$0.42 & 0.009$\pm$0.012 & 0.014$\pm$0.009 & 11.82$\pm$0.44 & 11.99$\pm$0.43 & 277$\pm$35  & 124$\pm$69  & 169 $\pm$335 & 162 $\pm$98    \\
0.49$\leq$$z_{cl}$$<$0.57                       & 26516 & 0.53$_{0.50}^{0.55}$ & 0.030$\pm$0.008 & 0.011$\pm$0.004 & 11.67$\pm$0.41 & 11.22$\pm$0.42 & 0.021$\pm$0.008 & 0.011$\pm$0.005 & 12.16$\pm$0.42 & 11.88$\pm$0.42 & 314$\pm$74  & 344$\pm$102 & -150$\pm$180 & -258$\pm$132   \\
$z_{cl}$$>$0.57                                 & 26523 & 0.62$_{0.58}^{0.67}$ & 0.045$\pm$0.007 & 0.010$\pm$0.004 & 11.85$\pm$0.40 & 11.20$\pm$0.42 & 0.011$\pm$0.006 & 0.006$\pm$0.004 & 11.90$\pm$0.42 & 11.60$\pm$0.43 & 348$\pm$41  & 379$\pm$92  & -38 $\pm$113 & 207 $\pm$90    \\\\

\hline \\ 

\multicolumn{1}{c}{Bin} & $N_{\rm pairs}$ & M$_{500}$ & \multicolumn{2}{c}{REW(\MgII) in \AA}  & \multicolumn{2}{c}{$\log_{10}N$(\MgII)/cm$^{-2}$}  & \multicolumn{2}{c}{REW(\FeII) in \AA} & \multicolumn{2}{c}{$\log_{10}N$(\FeII)/cm$^{-2}$} & \multicolumn{2}{c}{$\sigma_{v}$ (\kms)} &  \multicolumn{2}{c}{$V_{0}$ (\kms)} \\
               &         &                &  Mean & Median & Mean & Median & Mean & Median & Mean & Median & Mean & Median & Mean & Median  \\
 (1)           &  (2)    &  (3)          &   (4)        &    (5)           &   (6)         &   (7)    &  (8)   & (9) & (10)   & (11) & (12) & (13) & (14) & (15)      \\ \hline \hline

M$_{500}<$1.32           & 26476 & 1.13$_{0.84}^{1.26}$ & 0.034$\pm$0.008 & 0.011$\pm$0.005 & 11.73$\pm$0.41 & 11.24$\pm$0.42 & 0.008$\pm$0.009 & 0.007$\pm$0.005 & 11.73$\pm$0.44 & 11.71$\pm$0.43 & 269$\pm$28  & 260$\pm$49  & -93 $\pm$223 & -73 $\pm$132   \\
1.32$\leq$M$_{500}<$1.89 & 26512 & 1.56$_{1.39}^{1.77}$ & 0.034$\pm$0.007 & 0.011$\pm$0.004 & 11.73$\pm$0.41 & 11.22$\pm$0.42 & 0.016$\pm$0.008 & 0.005$\pm$0.004 & 12.04$\pm$0.42 & 11.58$\pm$0.43 & 407$\pm$42  & 277$\pm$142 & 219 $\pm$156 & 287 $\pm$209   \\
M$_{500}>$1.89           & 26497 & 2.52$_{2.04}^{3.72}$ & 0.038$\pm$0.007 & 0.010$\pm$0.004 & 11.78$\pm$0.41 & 11.19$\pm$0.42 & 0.024$\pm$0.007 & 0.013$\pm$0.005 & 12.22$\pm$0.42 & 11.96$\pm$0.42 & 441$\pm$119 & 158$\pm$33  & 141 $\pm$137 & 108 $\pm$167   \\\\

\hline \\ 
\multicolumn{1}{c}{Bin} & $N_{\rm pairs}$ & $\rho_{cl}$ & \multicolumn{2}{c}{REW(\MgII) in \AA}  & \multicolumn{2}{c}{$\log_{10}N$(\MgII)/cm$^{-2}$}  & \multicolumn{2}{c}{REW(\FeII) in \AA} & \multicolumn{2}{c}{$\log_{10}N$(\FeII)/cm$^{-2}$} & \multicolumn{2}{c}{$\sigma_{v}$ (\kms)} &  \multicolumn{2}{c}{$V_{0}$ (\kms)} \\
               &         &                &  Mean & Median & Mean & Median & Mean & Median & Mean & Median & Mean & Median & Mean & Median  \\
 (1)           &  (2)    &  (3)          &   (4)        &    (5)           &   (6)         &   (7)    &  (8)   & (9) & (10)   & (11) & (12) & (13) & (14) & (15)      \\ \hline \hline

$\rho_{cl}$$<$2.0                          & 26495 & 1.48$_{1.01}^{1.85}$ & 0.042$\pm$0.008 & 0.015$\pm$0.005 & 11.82$\pm$0.41 & 11.37$\pm$0.42 & 0.012$\pm$0.010 & 0.007$\pm$0.005 & 11.93$\pm$0.43 & 11.67$\pm$0.43 & 336$\pm$43  & 323$\pm$64  & 58  $\pm$208 & 48  $\pm$114   \\
2.0$\leq$$\rho_{cl}$$<$2.76                & 26494 & 2.40$_{2.13}^{2.65}$ & 0.038$\pm$0.007 & 0.010$\pm$0.004 & 11.78$\pm$0.41 & 11.22$\pm$0.42 & 0.022$\pm$0.008 & 0.010$\pm$0.005 & 12.18$\pm$0.42 & 11.85$\pm$0.42 & 438$\pm$87  & 406$\pm$108 & -68 $\pm$173 & 225 $\pm$96    \\
$\rho_{cl}$$>$2.76                         & 26496 & 3.14$_{2.88}^{3.66}$ & 0.028$\pm$0.008 & 0.008$\pm$0.004 & 11.64$\pm$0.41 & 11.08$\pm$0.42 & 0.013$\pm$0.009 & 0.009$\pm$0.005 & 11.95$\pm$0.43 & 11.80$\pm$0.43 & 244$\pm$29  & 178$\pm$43  & 144 $\pm$222 & 155 $\pm$131   \\\\

\hline \\ 
\multicolumn{1}{c}{Bin} & $N_{\rm pairs}$ & $\rho_{cl}$/r$_{500}$ & \multicolumn{2}{c}{REW(\MgII) in \AA}  & \multicolumn{2}{c}{$\log_{10}N$(\MgII)/cm$^{-2}$}  & \multicolumn{2}{c}{REW(\FeII) in \AA} & \multicolumn{2}{c}{$\log_{10}N$(\FeII)/cm$^{-2}$} & \multicolumn{2}{c}{$\sigma_{v}$ (\kms)} &  \multicolumn{2}{c}{$V_{0}$ (\kms)} \\
               &         &                &  Mean & Median & Mean & Median & Mean & Median & Mean & Median & Mean & Median & Mean & Median  \\
 (1)           &  (2)    &  (3)          &   (4)        &    (5)           &   (6)         &   (7)    &  (8)   & (9) & (10)   & (11) & (12) & (13) & (14) & (15)      \\ \hline \hline

$\rho_{cl}/R_{500}$$<$3.0                       & 26495 & 2.23$_{1.52}^{2.78}$ & 0.037$\pm$0.008 & 0.014$\pm$0.004 & 11.77$\pm$0.41 & 11.35$\pm$0.41 & 0.012$\pm$0.009 & 0.008$\pm$0.005 & 11.94$\pm$0.43 & 11.73$\pm$0.43 & 404$\pm$50  & 384$\pm$86  & 171 $\pm$145 & 125 $\pm$114   \\
3.0$\leq$$\rho_{cl}/R_{500}$$<$4.12             & 26494 & 3.61$_{3.21}^{3.96}$ & 0.045$\pm$0.008 & 0.013$\pm$0.005 & 11.85$\pm$0.41 & 11.30$\pm$0.42 & 0.016$\pm$0.009 & 0.007$\pm$0.005 & 12.05$\pm$0.43 & 11.69$\pm$0.43 & 336$\pm$47  & 296$\pm$133 & -65 $\pm$204 & -29 $\pm$236   \\
$\rho_{cl}/R_{500}$$>$4.12                      & 26496 & 4.58$_{4.28}^{4.87}$ & 0.024$\pm$0.007 & 0.008$\pm$0.005 & 11.57$\pm$0.42 & 11.09$\pm$0.43 & 0.018$\pm$0.009 & 0.010$\pm$0.005 & 12.09$\pm$0.42 & 11.86$\pm$0.42 & 294$\pm$46  & 204$\pm$103 & 114 $\pm$198 & 145 $\pm$204   \\\\
\hline \\ 
\multicolumn{16}{l}{Notes -- (1) Bin range. (2) Number of quasar-cluster pairs in that bin. (3) median value of the binning parameter in that bin. The upper and lower values in (3) indicate the 16 and 84 percentiles}\\
\multicolumn{16}{l}{of the parameter distribution. (4) \& (5) Rest-frame EWs of \mgii line measured between 2785$-$2815~\AA~ from the mean and median stack spectra, respectively. (6) \& (7) Column density estimated}\\

\multicolumn{16}{l}{using the \MgII\ REWs in (4) and (5), respectively, assuming lines are in the linear part of the curve-of-growth. (8) \& (9) Rest-frame EWs of \feii line measured between  2589$-$2611~\AA~ from the }\\

\multicolumn{16}{l}{mean and median stack profiles, respectively. (10) \& (11) Same as (6) \& (7) but for \feii line. (12) \& (13) Line widths calculated from three-component Gaussian fits to the \feiia and \mgiiab lines}\\
\multicolumn{16}{l}{for the mean and median stacks, respectively. We tied the line widths of the \mgii and \feii lines. (14) \& (15) Line centroids obtained from the fitting for the mean and median stack spectra,}\\ \multicolumn{16}{l}{respectively.}\\
\end{tabular}
\end{threeparttable}
}
\end{center}

\end{sidewaystable}

\end{document}